\documentclass[english]{article}
\usepackage[T1]{fontenc}
\usepackage[latin9]{inputenc}
\usepackage{geometry}
\usepackage{booktabs}
\usepackage{units}
\usepackage{amsmath}
\usepackage{amsthm}
\usepackage{amssymb}
\usepackage{graphicx}
\usepackage{soul}
\usepackage{color}
\newcommand{\comment}[1]{}
\usepackage{algorithm}
\usepackage{algpseudocode}

\newcommand{\cI}{\mathcal{I}}

\makeatletter

\providecommand{\tabularnewline}{\\}

\newcommand{\address}[1]{
	\par {\raggedright #1
	\vspace{1.4em}
	\noindent\par}
}

\numberwithin{figure}{section}
\theoremstyle{plain}

\@ifundefined{showcaptionsetup}{}{%
 \PassOptionsToPackage{caption=false}{subfig}}
\usepackage{subfig}
\makeatother

\usepackage{babel}

\begin{document}
\title{A Data-Driven Statistical-Stochastic Surrogate Modeling Strategy for Complex Nonlinear Non-stationary Dynamics}
\author{Di Qi\textsuperscript{a} and John Harlim\textsuperscript{b}}


\maketitle
\address{\textsuperscript{a }Department of Mathematics, Purdue University, West Lafayette, IN 47907, USA. Email: \texttt{qidi@purdue.edu}}
\address{\textsuperscript{b }Department of Mathematics, Department of Meteorology and Atmospheric Science, Institute for Computational and Data Sciences, The Pennsylvania State University, University Park, PA 16802. Email: \texttt{jharlim@psu.edu}
}

\begin{abstract}
We propose a statistical-stochastic surrogate modeling approach to predict the response of the mean and variance statistics under various initial conditions and external forcing perturbations. The proposed modeling framework extends the purely statistical modeling approach that is practically limited to the homogeneous statistical regime for high-dimensional state variables. The new closure system allows one to overcome several practical issues that emerge in the non-homogeneous statistical regimes. First, the proposed ensemble modeling that couples the mean statistics and stochastic fluctuations naturally produces positive-definite covariance matrix estimation, which is a challenging issue that hampers the purely statistical modeling approaches. Second, the proposed closure model, which embeds a non-Markovian neural-network model for the unresolved fluxes such that the variance of the dynamics is consistent, overcomes the inherent instability of the stochastic fluctuation dynamics. Effectively, the proposed framework extends the classical stochastic parametric modeling paradigm for the unresolved dynamics to a semi-parametric parameterization with a residual Long-Short-Term-Memory neural network architecture. Third, based on empirical information metric, we provide an efficient and effective training procedure by fitting a loss function that measures the differences between response statistics. Supporting numerical examples are provided with the Lorenz-96 model, a system of ODEs that admits the characteristic of chaotic dynamics with both homogeneous and inhomogeneous statistical regimes. In the latter case, we will see the effectiveness of the statistical prediction even though the resolved Fourier modes corresponding to the leading mean energy and variance spectra do not coincide. 
\end{abstract}

\section{Introduction and background\label{sec:Introduction-and-background}}

One of the key challenges in uncertainty quantification of dynamical systems \cite{leith1975climate,majda2018strategies,majda2019linear} and data assimilation \cite{majda2012filtering,reich2015probabilistic,h:15} is to construct a surrogate model that allows one to accurately and efficiently predict the evolution of the low-order statistics under perturbation of model parameters (e.g., additional forces) and initial conditions. Computationally, how uncertainty propagates in dynamical systems is usually characterized by understanding how the second-order statistics change when the system's initial conditions and/or parameters are perturbed. Under mild perturbations and appropriate mathematical conditions, this problem has also been studied in non-equilibrium statistical mechanics (see e.g. Chapter 7 of \cite{zwanzig2001nonequilibrium}). The time evolution of mean and covariance statistics is also essential in data assimilation. In this application, most algorithms (such as Kalman filtering and its variant) construct conditional mean and covariance statistics by a Bayesian formula that updates these low-order statistics to account for the newly measured observations. Subsequently, the resulting conditional statistics are fed into the dynamical model as initial conditions for predicting the state with the mean and uncertainty characterized by the covariance statistics. 

An expensive method to compute these statistics is to employ a Monte-Carlo simulation. This approach involves solving an ensemble of solutions for the dynamical systems and using these solutions to empirically estimate the statistics of interest. In the data assimilation context, such an idea has been realized by the well-known Ensemble Kalman Filter algorithm \cite{evensen2003ensemble}. With the Monte-Carlo approach, the computational cost is determined by the complexity of integrating the dynamical system multiplying the ensemble size. One known issue with such an approach is that the ensemble size required to maintain the desired accuracy will grow exponentially as a function of the state-spaced dimension. This issue poses major computational challenge when online statistical predictions under new initial conditions and forces are needed in uncertainty quantification and data assimilation applications. The approach adopted in this paper is to construct a reduced-order model to accurately predict the evolution of the low-order statistics, where an offline machine learning algorithm is employed to emulate the feedback from higher-order statistical moments. 

While the proposed formulation here is applicable to any complex spatially extended nonlinear dynamical systems, it is also closely related to a long-standing moment closure problem of turbulent dynamical systems that has been widely studied in many fields of science and engineering \cite{lesieur1987turbulence,SapsisMajda2013,qi2016low}. This serendipity motivates us to present the formulation on a specific class of nonlinear dynamical system, where the moment interactions are induced by a bilinear quadratic form that is typically inherited from a discretization (or spectral projection) of nonlinear advection in fluid dynamical models, such as in the Navier-Stokes and Burger's equations, and the spatiotemporally chaotic  Kuramoto-Sivashinsky equation. In any nonlinear systems, the moment dynamics is not closed in the sense that the dynamical equation for each moment depends on the higher-order moments in addition to the lower-order statistics. This inherent hierarchical structure poses some practical issues especially if one is interested to resolve at least the first- and second-order moments. Particularly, for a system with $N$-dimensional state space variables, while the evolution of the first-order moment is represented by a system of $N$-dimensional differential equations, the evolution of the second-order moment is represented by an $N\times N$ matrix-valued differential equations that further depend on components of the unresolved third-order moments of size $N\times N\times N$. 
 
For the first-order moment closure of turbulent dynamics, machine learning has been used to approximate the unresolved subgrid scale terms \cite{gamahara2017searching,singh2017machine,maulik2019subgrid}). In fact, for the second-order moment closure problem, machine learning with Long-Short-Term-Memory (LSTM) architecture has been proposed \cite{qi2022machine} to emulate the feedback from the third-order moments. 
The approach in \cite{qi2022machine}, unfortunately, is restricted to spatially homogeneous statistics. When the statistics are spatially homogeneous, the mean and variance dynamics can be represented by a $\left(1+N\right)$-dimensional system, consisting of the one-dimensional mean variable and $N$-dimensional variance components since the non-diagonal entries in the covariance matrix are all zeros. Such a reduced representation, unfortunately, is invalid in non-homogeneous systems. The work in this paper is to extend the machine learning approach in \cite{qi2022machine} to non-homogeneous statistics. In this regime, several issues emerge. Beyond the practical issue of resolving the $N\times N$ non-diagonal covariance matrix, constructing a machine learning model for the feedback from the third-order moments that preserves the positive-definite covariance remains a challenging task, especially when the covariance dynamical equation is conditionally unstable due to the quasilinear coupling with the mean state.

To overcome this issue, we propose a reduced-order system of differential equations for the leading-order mean state and stochastic fluctuations. In this fully coupled statistical-stochastic system, the reduced-order mean dynamics depend on the covariance matrix that is empirically estimated using the ensemble prediction of the fluctuation terms and the resolved mean state determines the linear stability of the stochastic fluctuation dynamics. Denoting the resolved state space dimension by  $K$, where $K<N$, the dynamical equation of the proposed statistical-stochastic model has the complexity of order $K(1+ M)$, accounting for the $K$-dimensional vector for the mean and the $M$ ensemble members of $K$-dimensional fluctuation equations. While the system can be moderately high-dimensional, especially if the ensemble size $M > K$, the unresolved processes to be modeled in this formulation are only $K$-dimensional, accounting for the modeling error in the $K$-dimensional reduced-order mean dynamics and the unresolved fluxes in the $K$-dimensional fluctuation dynamics. 

For a scalable and accurate statistical prediction on this systematic modeling framework, two issues need to be addressed. First, internal instability as a common feature in chaotic dynamics often leads to unstable dynamic and fast divergence of the solution when the data-driven surrogate model is chosen from an arbitrary hypothesis space (such as the neural-network models) without any dynamical constraints.
To overcome this issue, we will impose dynamical constraints through a semi-parametric framework with consistent variances, embedding the neural-network model on a parametric modeling framework proposed in \cite{qi2022machine, majda2018strategies}. Second, while the ensemble structure in the proposed statistical-stochastic system is natural for ensemble prediction and data assimilation, the computational cost of this fully coupled system {$K(1+ M)$} is often too high for accurate learning. In our numerical example, the state-space dimension is of order 1000 for a {$K= 14$} dimensional reduced-order state variable. In this case, the standard learning procedure of fitting an empirical loss function that compares the trajectories of these states \cite{HJLY:19} may not be numerically feasible since it requires a very large training data set resulting in an enormously expensive computational cost . In the Appendix of this paper, we documented that applying such a learning procedure with a generic neural-network model to identify the unresolved fluxes in the fluctuation dynamics leads to overfitting. To overcome this issue, we will consider an empirical information metric as a loss function \cite{chen2022physics} that compares the response mean and variance statistics. We will show that fitting to the statistical responses corresponding to the same trajectories that led to overfitting in the standard procedure produces accurate response statistical prediction subject to new initial conditions and external forces, both in homogeneous and inhomogeneous statistical regimes. 

{ As we already mentioned above, while the approach can be implemented on any spatially extended chaotic dynamical systems, since the approach is closely related to the moment closure problem, we will present our approach on nonlinear systems with a bilinear form.} Specifically, we will examine the stochastic-statistical formulation of the Lorenz-96 model \cite{lorenz:96} which admits the characteristic of the chaotic dynamical systems and can be adjusted to generate non-trivial inhomogeneous statistics. First, we should point out that the variance spectrum in this system decays slowly (relative to the Kolmogorov decay in classical turbulence theory) and the corresponding Fourier modes with large variance spectrum are all unstable. When spatially large-scale disturbances are injected into the system, while they excite the entire mean energy spectrum and variance spectrum, the changes in the mean energy spectrum are significantly noticeable in the Fourier modes corresponding to the lower variance spectrum. The mismatch between modes that have a large mean energy spectrum and those that have a large variance spectrum makes this system ideal for testing the proposed reduced-order statistical-stochastic framework. Particularly, this test model would allow us to understand to which extent the reduced ordering can be employed and whether the internal instability in this system can be overcome with the proposed modeling framework.      

The remainder of this paper is organized as follows. In Section~\ref{sec:General-mathematical-formulation}, we discuss the general statistical-stochastic closure modeling framework of turbulent dynamical systems. In Section~\ref{sec:Machine-learning-strategies}, we discuss the proposed machine learning strategy on a concrete example, the Lorenz-96 model. In Section~\ref{sec:Predicting-leading-order}, we discuss the training configuration and present numerical results for the proposed closure framework, both on homogeneous and inhomogeneous statistical regimes. In Section~\ref{sec:Summarizing-discussions}, we close the paper with a summary. As mentioned in the above discussion, we include an example demonstrating the difficulty of attaining accurate trajectory prediction on the test dataset using the standard machine learning procedure in \ref{appen:Predicting-leading-order-statist}, which motivates this work.

\section{General mathematical formulation for {complex} systems with uncertainty\label{sec:General-mathematical-formulation}}

In this section, we give a quick overview of a general moment closure formulation for a class of complex nonlinear systems that is common in natural and engineering problems and formulate an efficient machine learning reduced-order model. One representative feature that makes the moment closure problem challenging in such complex systems is the nonlinear energy-conserving interaction that transports energy across scales. The general formulation of the turbulent dynamical systems can be characterized by the canonical equations of the state variable $\mathbf{u}\in\mathbb{R}^{N}$ in a high-dimensional phase space,
\begin{equation}
\frac{d\mathbf{u}}{dt}=\left(\mathcal{L}+\cal{D}\right)\mathbf{u}+B\left(\mathbf{u},\mathbf{u}\right)+\mathbf{F}+\sigma\dot{\mathbf{W}}.\label{eq:abs_formu}
\end{equation}
On the right hand side of the equation (\ref{eq:abs_formu}), the
first two components, $\left(\mathcal{L}+\cal{D}\right)\mathbf{u}$, represent linear
dispersion and dissipation effects, where $\mathcal{L}^{*}=-\mathcal{L}$ is
an energy-conserving skew-symmetric operator for dispersive effects;
and $\mathcal{D}<0$ is a negative definite operator for dissipations. The nonlinear
effect in the dynamical system is introduced through a quadratic form,
$B\left(\mathbf{u},\mathbf{u}\right)$ \cite{majda2018strategies} { that arise in a discretization of the nonlinear advection in fluid mechanics.} Besides, the system is usually subject to time-dependent external forcing effects that are decomposed into a deterministic component, $\mathbf{F}\left(t\right)$,
and a stochastic component represented by a Gaussian random process,
$\sigma\left(t\right)\dot{\mathbf{W}}\left(t;\omega\right)$.
It needs to be emphasized that in many situations $\mathbf{F}$ might
be spatially inhomogeneous, and thus, introduce anisotropic structures into the system.

One way to characterize the effect of internal instabilities and the uncertainties from the initial state and forcing in the turbulent system (\ref{eq:abs_formu}) is through  a statistical description for the time evolution of the moment of the state variable $\mathbf{u}$. While in principle the dynamical equations of the statistical moments follow the backward-Kolmogorov PDE (which is the $L^2$ adjoint of the Fokker-Planck equation that characterizes the evolution of the density function $p(\mathbf{u},t)$), it remains challenging to computationally solve such a PDE, especially when state space dimension, $N$, is large.
The Monte-Carlo approach \cite{leutbecher2008ensemble, toth1997ensemble}, which uses an ensemble of solutions of \eqref{eq:abs_formu} subjected to initial and forcing perturbations, provides an alternative means to quantify the essential statistics that quantify the  uncertainties through empirical ensemble averages.

\subsection{The exact formulation for statistical mean and stochastic fluctuation interactions}\label{sec2.1}

Despite its simplicity, a direct ensemble forecast obtained from integrating the original
model (\ref{eq:abs_formu}) has several difficulties in accurately
recovering the key model statistics in a high dimensional space.
First, the ensemble size required to maintain the accuracy will grow exponentially in direct ensemble simulation of the full model as the dimension of the system increases. This requirement may not be computationally desirable, especially when online predictions under new initial conditions and forces are needed, and subsequently motivates the need for reduced-order modeling. On the other hand, turbulent systems often contain strong internal instability and mixed {spatio-temporal} structures. These features pose some computational challenges for developing effective reduced-order models directly under the original model formulation, especially when the reference dynamical system is nonlinear and non-Gaussian. 

To address these difficulties, we introduce a statistical-stochastic
decomposition of the model state $\mathbf{u}$, so that the mean-fluctuation interactions can be identified. Efficient model reduction
strategies will be proposed where data-driven components can be introduced naturally to account for the unresolved fluctuation interactions. To achieve this, we view the model state $\mathbf{u}$ as a random field and project it onto the composition of a statistical mean and stochastic fluctuations in a finite-dimensional representation under a suitable orthonormal basis $\left\{ \mathbf{e}_{i}\right\} _{i=1}^{N}$ as,
\begin{equation}
\mathbf{u}\left(t;\omega\right)=\bar{\mathbf{u}}(t)+\mathbf{u}^{\prime}(t;\omega)=\bar{\mathbf{u}}\left(t\right)+\sum_{i=1}^{N}Z_{i}\left(t;\omega\right)\mathbf{e}_{i},\label{eq:spec_expansion}
\end{equation}
where $\bar{\mathbf{u}}\left(t\right)=\left\langle \mathbf{u}\left(t\right)\right\rangle $
 (here and after, we use $\left\langle \cdot\right\rangle $ to denote the statistical expectation about the PDF $p\left(\mathbf{u},t\right)$), represents the statistical expectation of the model state, i.e. the mean field; and $\left\{ Z_{i}\left(t;\omega\right)\right\} $
as the mean-zero stochastic coefficients measuring the uncertainty in
fluctuation processes $\mathbf{u}^{\prime}$ along each \st{eigenmode} direction 
 $\mathbf{e}_{i}$. The statistical uncertainty among the
fluctuation modes can be characterized by the covariance between the
stochastic modes. 

By taking the statistical (ensemble) average over the original equation
(\ref{eq:abs_formu}) and using the mean-fluctuation decomposition
(\ref{eq:spec_expansion}), the \emph{evolution equation of the statistical
mean state}\textbf{ $\bar{\mathbf{u}}$} is given by the following
dynamical equation,
\begin{equation}
\frac{d\bar{\mathbf{u}}}{dt}=\left(\mathcal{L}+\mathcal{D}\right)\bar{\mathbf{u}}+B\left(\bar{\mathbf{u}},\bar{\mathbf{u}}\right)+\sum_{i,j=1}^{N}R_{ij}B\left(\mathbf{e}_{i},\mathbf{e}_{j}\right)+\mathbf{F},\label{eq:mean_dyn}
\end{equation}
where $R:=\left\langle \mathbf{Z}\mathbf{Z}^{*}\right\rangle $ denotes the
second-order covariance matrix of the stochastic coefficients $\mathbf{Z}=\left\{ Z_{i}\right\} _{i=1}^{N}$.
The term $B\left(\bar{\mathbf{u}},\bar{\mathbf{u}}\right)$ represents
the nonlinear interactions between the mean state, and $R_{ij}B\left(\mathbf{e}_{i},\mathbf{e}_{j}\right)$
is the higher-order feedback from the fluctuation modes to the mean
state dynamics. Next, by projecting the above equation (\ref{eq:abs_formu}) to each orthonormal
basis element $\mathbf{e}_{i}$ we obtain the \emph{evolution equation
for the stochastic fluctuation coefficients},
\begin{equation}
\frac{dZ_{i}}{dt}=\sum_{j=1}^{N}A_{ij}\left(\bar{\mathbf{u}}\right)Z_{j}+\sum_{m,n=1}^{N}\gamma_{imn}\left(Z_{m}Z_{n}^{*}-R_{mn}\right)+\sigma\dot{\mathbf{W}}\cdot\mathbf{e}_{i},\label{eq:coeff_dyn}
\end{equation}
where $A_{ij}\left(\bar{\mathbf{u}}\right)=\left[\left(\mathcal{L}+\mathcal{D}\right)\mathbf{e}_{j}+B\left(\bar{\mathbf{u}},\mathbf{e}_{j}\right)+B\left(\mathbf{e}_{j},\bar{\mathbf{u}}\right)\right]\cdot\mathbf{e}_{i}$
characterizes the quasilinear coupling between the mean state $\bar{\mathbf{u}}$
and the fluctuations $\mathbf{u}^{\prime}=\sum_{i}Z_{i}\mathbf{e}_{i}$.
The interactions between the fluctuation modes of different scales
are summarized in the second term on the right hand side of (\ref{eq:coeff_dyn})
with the coupling coefficient $\gamma_{imn}=B\left(\mathbf{e}_{m},\mathbf{e}_{n}\right)\cdot\mathbf{e}_{i}$.
Alternatively, from the stochastic equation (\ref{eq:coeff_dyn})
we directly obtain the exact \emph{evolution equation of the covariance
matrix} $R$, 
\begin{equation}
\frac{dR}{dt}=A\left(\bar{\mathbf{u}}\right)R+RA^{*}\left(\bar{\mathbf{u}}\right)+Q_{F}+Q_{\sigma},\label{eq:cov_dyn}
\end{equation}
where $A\left(\bar{\mathbf{u}}\right)$ is the same quasilinear operator
from (\ref{eq:coeff_dyn}) {containing instability represented by its positive eigenvalues, while $Q_{F}$ is the nonlinear energy flux, which includes all the third moments $\left\langle Z_{m}Z_{n}Z_{i}\right\rangle$ feedback to balance the the unstable linear growth.} {The term $Q_{\sigma,kl}=\sum_{m}\left(\mathbf{e}_{k}\cdot\sigma_{m}\right)\left(\sigma_{m}\cdot\mathbf{e}_{l}\right)$}
is the contribution from the unresolved white noise forcing. Detailed expression for the equation (\ref{eq:cov_dyn}) can be found in \cite{majda2018strategies}. 

As a further remark on this mean-fluctuation formulation of the original system, we could use either the stochastic equation (\ref{eq:coeff_dyn})
or the equivalent statistical covariance equation (\ref{eq:cov_dyn}) to model the uncertainty in each fluctuation mode $\mathbf{e}_{i}$. In fact, a data-driven statistical closure model combining (\ref{eq:mean_dyn}) and
(\ref{eq:cov_dyn}) has been developed in \cite{qi2022machine} to
effectively capture the leading-order statistical responses in mean
and variance of homogeneous turbulent dynamics. On the other hand, the statistical-stochastic formulation using (\ref{eq:mean_dyn}) and (\ref{eq:coeff_dyn}) enjoys the advantage of more flexibility to run ensemble forecasts for both uncertainty quantification and data assimilation, { and compute statistical quantities other than the covariance.} In addition, this statistical-stochastic model can naturally estimate inhomogeneous statistics and avoids the main issue with the purely statistical formulation in \eqref{eq:mean_dyn}, \eqref{eq:cov_dyn}  in preserving the positive-definite covariance estimation.

\subsection{A generic statistical-stochastic closure model for mean and variance statistics}

Now, we present the main idea in the efficient combined statistical-stochastic
model to effectively capture the central statistical features. To effectively reduce the computational cost in finding
the solution of high dimensional phase space, we introduce a proper
low wavenumber truncation so that only the most important leading
modes in the subset $\cI$ ({for example, the subset $\cI$ can be taken to include the most energetic modes $\mathbf{e}_i$ in the projection \eqref{eq:spec_expansion} with the largest mean energy and/or variances}) are resolved, that is, 
\begin{equation}
\mathbf{u}^{\cI}=\bar{\mathbf{u}}^{\cI}+\sum_{i\in\cI}Z_{i}\mathbf{e}_{i},\label{eq:state_decomp}
\end{equation}
where $\bar{\mathbf{u}}^{\cI}=\mathrm{Pr}_{\cI}\bar{\mathbf{u}}=\sum_{i\in\cI}\bar{u}_{i}\mathbf{e}_{i}$ is a low-dimensional representation of the mean state and $Z_i$ denotes the stochastic coefficients corresponding to the low-dimensional subset of the full state space $\left|\cI \right| = { K}\ll N$. Inspecting the coupling terms in the true dynamics (\ref{eq:mean_dyn}) and (\ref{eq:coeff_dyn}),
several difficulties will emerge for accurate modeling of the detailed
coupling mechanisms in the constrained reduced-order representation
(\ref{eq:state_decomp}). First, the high-order nonlinear coupling terms
in the mean and fluctuation equations consist of a {multiscale interaction
of modes along the entire spectrum}, while we only have access to a
 subset $\cI$ of the resolved mean and fluctuation modes. Second, inherent instability in the fluctuation modes $Z_{i}$ due to the quasilinear coupling with $\mathbf{\bar{u}}$ through $A_{ij}\left(\bar{\mathbf{u}}\right)$ poses challenge { in constructing a stable dynamical model that can accurately predict the statistical} responses to various perturbations. Third, the fluctuation components $Z_{i}$ are stochastic processes coupled to the statistical mean equation, making direct modeling of the random trajectories {very expensive. }

\subsubsection{Effective closure equations for the mean and fluctuations}\label{sec2.2.1}

First, we introduce the \emph{reduced statistical mean equation} by
projecting the full equation (\ref{eq:mean_dyn}) to the resolved
low-dimensional subspace 
\begin{equation}
\frac{d\bar{\mathbf{u}}^{\cI}}{dt}=\left(\mathcal{L}+\mathcal{D}\right)\bar{\mathbf{u}}^{\cI}+\mathrm{Pr}_{\cI}B\left(\bar{\mathbf{u}}^{\cI},\bar{\mathbf{u}}^{\cI}\right)+\sum_{i,j\in\cI}R_{ij}\mathrm{Pr}_{\cI} B\left(\mathbf{e}_{i},\mathbf{e}_{j}\right)+\mathbf{F}^{\cI}+\boldsymbol{\Theta}^{m}.\label{eq:dyn_mean_clos}
\end{equation}
In the above equation, only the projected dynamics in the reduced
subspace are resolved. { Here, the unresolved mean feedback that we denoted as $\boldsymbol{\Theta}^{m}$ accounts for the residual (or truncation error) induced by the projection, namely the difference between the right hand side of the full model in \eqref{eq:mean_dyn} and the first three resolved components in the right-hand-side of \eqref{eq:dyn_mean_clos}.} Various statistical closure strategies have been developed \cite{majda2018strategies,majda2016introduction} using the parametric approximation of the unresolved structures.  In this paper, we aim to design a machine learning scheme to identify this unresolved process directly from data.

Second, we consider the stochastic closure for the fluctuation equation
(\ref{eq:coeff_dyn}). Again we concentrate on modes in the subset $\cI$ as in (\ref{eq:state_decomp}). 
Let $\mathbf{Z}^{\cI}=\mathrm{Pr}_{\cI}\mathbf{Z}=\big\{ Z_{i}\big\}_{i\in\cI}$
be the vector of the resolved fluctuation modes. Similar to the statistical mean closure (\ref{eq:dyn_mean_clos}), we propose to construct a projected dynamical model for the resolved modes and learn the unresolved feedback with a properly designed data-driven model. Specifically, the resulting
\emph{reduced-order fluctuation equation} for the stochastic coefficients
$\mathbf{Z}$ becomes 
\begin{equation}
\frac{d\mathbf{Z}^{\cI}}{dt}=A\left(\bar{\mathbf{u}}^{\cI}\right)\mathbf{Z}^{\cI}+\sigma\dot{\mathbf{W}}^{\cI}\:+\boldsymbol{\Theta}^{v},\label{eq:dyn_coeff_clos}
\end{equation}
{ where $\boldsymbol{\Theta}^{v}$ denotes the residual induced by the projection, namely the difference between the right hand side of the full model in \eqref{eq:cov_dyn} and the first two resolved components in the right-hand-side of \eqref{eq:dyn_coeff_clos}.}

Notice that the quasilinear coefficient $A_{ij}\left(\bar{\mathbf{u}}\right)=\left[\left(\mathcal{L}+\mathcal{D}\right)\mathbf{e}_{j}+B\left(\bar{\mathbf{u}},\mathbf{e}_{j}\right)+B\left(\mathbf{e}_{j},\bar{\mathbf{u}}\right)\right]\cdot\mathbf{e}_{i}$
for $i,j\in\cI$ includes the mean-fluctuation interaction
leading to inherent internal instability for turbulent dynamics (that
is, with positive eigenvalues in $A\left(\bar{\mathbf{u}}\right)$).
{ While this linear instability is suppressed in the full model by the second term in \eqref{eq:coeff_dyn}, or equivalently by $Q_F$ in \eqref{eq:cov_dyn}, dynamical instability can occur in the reduced-order model, especially when the term $\boldsymbol{\Theta}^{v}$ is numerically approximated with an arbitrary class of hypothesis models. To construct a stable approximate dynamical equation that can suppress instability induced by the marginally stable dynamics in \eqref{eq:dyn_coeff_clos}}, we introduce a
more detailed parameterization for the unresolved process as
\begin{equation}
\boldsymbol{\Theta}^{v}=-D\mathbf{Z}^{\cI}+\Sigma \dot{\widetilde{\mathbf{W}}},\label{eq:flux_decomp}
\end{equation}
where the white noises $\dot{\widetilde{\mathbf{W}}}$ are independent to $\dot{\mathbf{W}}$. {Here, $D$ and $\Sigma$ are coefficient matrices to be approximated (see \eqref{damping_noise} below}).
In \eqref{eq:flux_decomp}, {the parameters $D$ and $\Sigma$ are introduced to play the equivalent role as the nonlinear flux term $Q_F$ in the corresponding statistical equation \eqref{eq:cov_dyn}: the parameter $D$ is introduced to act as the equivalent damping suppressing the unstable positive growth rate, while the parameter $\Sigma$ is to account for the energy source from the nonlinear exchange of energy.} The effective decomposition in
(\ref{eq:flux_decomp}) generalizes the idea in the statistical closure
model for nonlinear energy mechanism \cite{qi2022machine,majda2018strategies}. Here, effective parameters $D, \Sigma$ will be constructed by fitting to the consistent covariance statistics. {Applying It\^o's lemma to $f(\mathbf{Z}^{\cI}) = \frac{1}{2}\mathbf{Z}^{\cI}(\mathbf{Z}^{\cI})^*$ and taking expectation, we obtain, }
\[
\frac{dR^{\cI}}{dt}=A\left(\bar{\mathbf{u}}^{\cI}\right)R^{\cI}+R^{\cI}A^{*}\left(\bar{\mathbf{u}}^{\cI}\right)+Q_{\sigma}^{\cI}+Q_{F}^{\cI},
\]
where $R^{\cI}=\left\langle \mathbf{Z}^{\cI}(\mathbf{Z}^{\cI})^*\right\rangle $
is the covariance matrix of the resolved fluctuation modes, and
$Q_{F}^{\cI}$ is the nonlinear flux induced by the coupling from different
stochastic coefficients  and the truncation error. {When $\boldsymbol{\Theta}^{v}$ is parameterized by \eqref{eq:flux_decomp}, the It\^o's lemma for the flux parameterization is given by,
\begin{equation}
Q_F^{\cI} = -D R^{\cI} - R^{\cI}D^* + \Sigma\Sigma^*. \label{ideal_parm}    
\end{equation}
}
While such a choice of parameterization is ideal, it is difficult to numerically find $D$ and $\Sigma$ that satisfy \eqref{ideal_parm} as the covariance $R^{\cI}$ is a time-dependent variable. To avoid such a practical issue, we consider fitting to the stationary (equilibrium) statistics,
{
\begin{equation}
Q_{F}^{\mathcal{I}} \approx  - \tilde{D} R_{\mathrm{eq}}^{\cI} - R_{\mathrm{eq}}^{\cI}\tilde{D}^* + \tilde{\Sigma} \tilde{\Sigma}^* := \tilde{Q},\label{damping_noise_decomposition}
\end{equation}
where $R_{\mathrm{eq}}^{\cI}$ denotes the stationary covariance statistics of the resolved modes in $\cI$ that can be empirically estimated. In the above approximation, we introduced the notations $\tilde{Q}, \tilde{D},$ and $\tilde{\Sigma}$ to denote the approximate model (or parameterization). The above approximation in \eqref{damping_noise_decomposition} is to first enforce equilibrium consistency in the unperturbed case and then assume the identity is still valid for small perturbations. The approximation in \eqref{damping_noise_decomposition} becomes exact equality at the long-term limit guaranteeing the final convergence to the equilibrium covariance statistics of the unperturbed dynamics. With this approximation, we can decompose the approximate model into a positive and a negative definite component
$\tilde{Q} = (\tilde{Q})^+ - (\tilde{Q})^-$. Then the effective damping and noise
matrix can be approximated accordingly by fitting the negative and {positive-definite} components, respectively, as
\begin{equation}
D \approx \tilde{D}:=\frac{1}{2}(\tilde{Q})^{-}(R_{\mathrm{eq}}^{\cI})^{-1},\quad\Sigma\Sigma^* \approx\tilde{\Sigma}(\tilde{\Sigma})^* :=(\tilde{Q})^{+}.\label{damping_noise}
\end{equation}
The above approximation is based on the equivalent roles of the effective damping and noise as discussed above. We note that the equilibrium covariance matrix $R_{\mathrm{eq}}^{\cI}$ is nonsingular, and a well-conditioned matrix when the variance is not too small. This motivates the choice of modes in $\mathcal{I}$ with large variance energy spectra.} When the covariance matrix is diagonally dominant, we found that a further simplification can be made to avoid the computational complexity of realizing the matrix factorization above.
The general framework above, \eqref{eq:dyn_mean_clos}-\eqref{eq:flux_decomp} with the approximate coefficients {$\tilde{D}$ and $\tilde{\Sigma}$} in \eqref{damping_noise}, will be implemented on an explicit model next in Section~\ref{sec:Machine-learning-strategies} as a concrete example of the idea.

\subsubsection{An ensemble-based statistical and stochastic model}\label{sec2.2.2}

Finally, we need to couple the statistical mean equation (\ref{eq:dyn_mean_clos})
and the stochastic equation (\ref{eq:dyn_coeff_clos}) for the fluctuation
modes. The resolved mean state $\bar{\mathbf{u}}^{\cI}$ enters the
fluctuation equation (\ref{eq:dyn_coeff_clos}) through the quasilinear
term $A\left(\bar{\mathbf{u}}^{\cI}\right)$. Especially, it induces
positive growth rate among the unstable modes. Inversely, the statistical
mean equation (\ref{eq:dyn_mean_clos}) depends on the covariance feedback from the resolved modes $R^{\cI}$, which will be empirically estimated by a Monte-Carlo average over an ensemble of solutions of the fluctuation
equation (\ref{eq:dyn_coeff_clos}). Denoting the ensemble solutions of the fluctuation
equation (\ref{eq:dyn_coeff_clos}) as $\{\mathbf{Z}^{M,\left(i\right)}\}_{i=1,\ldots, M}$ the second-order moment can be estimated
empirically as, 
\begin{eqnarray}
R^{\cI}=\left\langle \mathbf{Z}^{\cI}(\mathbf{Z}^{\cI})^*\right\rangle \approx R^M :=\frac{1}{M-1}\sum_{i=1}^{M}\mathbf{Z}^{M,\left(i\right)}(\mathbf{Z}^{M,\left(i\right)})^*.\label{R^M_empirical}
\end{eqnarray}
With this empirical estimation, {we have the complete \emph{general reduced-order statistical-stochastic closure model} as},
\begin{eqnarray}
\begin{aligned}
\frac{d\bar{\mathbf{u}}^{M}}{dt}&=\left(\mathcal{L}+\mathcal{D}\right)\bar{\mathbf{u}}^{M}+\mathrm{Pr}_{\cI}B\left(\bar{\mathbf{u}}^{M},\bar{\mathbf{u}}^{M}\right)+\sum_{i,j\in\cI}R_{ij}^{M}\mathrm{Pr}_{\cI}B\left(\mathbf{e}_{i},\mathbf{e}_{j}\right)+\mathbf{F}^{\cI}+\boldsymbol{\Theta}^{m} \\
\frac{d\mathbf{Z}^{M}}{dt} &= A\left(\bar{\mathbf{u}}^{M}\right)\mathbf{Z}^{M}+\sigma\dot{\mathbf{W}}^{\cI}\; -D^M\mathbf{Z}^{M}+\Sigma^M \dot{\widetilde{\mathbf{W}}},\label{generalclosuremodel}
\end{aligned}
\end{eqnarray}
where {$\boldsymbol{\Theta}^{m} $ is the nonlinear mean feedback defined in \eqref{eq:dyn_mean_clos}. The coefficients $D^M$ and $\Sigma^M$ will be parameterized following the approximation in \eqref{damping_noise}, 
\begin{equation}
D^M:=\frac{1}{2}(Q^M)^{-}(R_{\mathrm{eq}}^{\cI})^{-1},\quad(\Sigma^M)(\Sigma^M)^*  :=(Q^M)^{+},\label{parameterizations} 
\end{equation}
where the model $Q^M = (Q^M)^{+} - (Q^M)^{-}$ approximates $Q_F^{\cI}$ is defined to follow \eqref{damping_noise_decomposition},
\[
Q^M := -D^M R^{\cI} - R^{\cI}(D^M)^* + \Sigma^M (\Sigma^M)^*.
\]  
The closure model \eqref{generalclosuremodel} with parameterization \eqref{parameterizations} provides a new formulation for the leading order statistics by combining the statistical mean equation with the stochastic fluctuation dynamics. Subsequently, the data-driven closure is adopted by fitting the standard Long Short Term Memory (LSTM) models to learn the
unresolved terms $\boldsymbol{\Theta}^{m}$ and $Q^M$.} In Section~\ref{sec:Machine-learning-strategies}, we will provide specific examples of \eqref{generalclosuremodel} induced by the moment closure of homogeneous and inhomogeneous turbulence dynamics.

\subsubsection{Empirical loss functions based on { an information metric}}\label{empirical_statistical_fitting}

It is important to design a suitable criterion for the loss function that reflects the appropriate quantity of interests, which are the response mean and variance statistics rather than the individual trajectory of the stochastic fluctuations {\cite{williams2006gaussian,goodfellow2016deep}}. In fact, we demonstrate numerically in \ref{appen:Predicting-leading-order-statist} that fitting the stochastic components of \eqref{generalclosuremodel} directly to the pathwise trajectory of \eqref{eq:dyn_coeff_clos} given the true mean $\bar{\mathbf{u}}^\cI$ often leads to overfitting, thus does not provide accurate statistical prediction when testing on new inputs. {In this context new inputs correspond to new initial values and forcing perturbations as in the numerical tests.} However, fitting to the mean and variance statistical responses corresponding to the same trajectory solutions (that lead to an overfitted model when trajectory is fitted) produces a closure model with accurate statistical predictions on new inputs. 

{A natural way to fit the statistics is to consider the information distance as it allows one to measure the errors between the probability distributions achieved from the empirical average of the ensemble simulations.} Particularly, we consider the following practical metric based on Kullback-Leibler (KL) divergence \cite{kullback1987letter,majda2005information} of two empirical measures induced by the response mean and variance statistics of the underlying dynamics in \eqref{eq:abs_formu} and the reduced-order model in \eqref{generalclosuremodel}, respectively.  {The KL divergence offers a balanced calibration between the statistical errors in the mean and variance.} Let $\delta \bar{\mathbf{u}}:=\mathbf{\bar{u}}_{\delta}-\mathbf{\bar{u}}_{\textup{eq}}$ and $\delta R: = R_\delta - R_{\textup{eq}}$ be the response mean and covariance statistics of the underlying dynamics in \eqref{eq:abs_formu} subject to additional damping and forcing of small $0<\delta \ll 1$ perturbation amplitudes in addition to the reference damping and forcing parameters. Here, the response statistics are defined as the differences between the time-dependent statistics subject to the additional damping and forces and the equilibrium statistics, $\mathbf{\bar{u}}_{\textup{eq}}$ and $R_{\textup{eq}}$, corresponding to the solutions with reference damping and forcing that can be empirically estimated offline by an ensemble simulation of the underlying system in \eqref{eq:abs_formu}. Analogously, we also define $\delta \bar{\mathbf{u}}^M:=\mathbf{\bar{u}}^{M}_\delta-\mathbf{\bar{u}}_{\textup{eq}}$ and $\delta R^M:=R^{M}_\delta - R_{\textup{eq}}$ as the corresponding statistical responses of the reduced-order model in \eqref{generalclosuremodel}, where $\mathcal{D}$ and $\mathbf{F}^\cI$ are perturbed by additional damping and forcing of amplitude $\delta$. Assuming that the perturbed distributions vary smoothly under parameter $\delta$ and denoting $\textup{diag}(R)$ as the diagonal matrix whose diagonal entries are $R_{k}$, the KL-divergence between Gaussian measures $\pi_\delta = \mathcal{N}(\mathbf{\bar{u}}_{\textup{eq}}+ \delta \bar{\mathbf{u}}, \textup{diag}(R_{\textup{eq}}+\delta R))$ and $\pi_\delta^M = \mathcal{N}(\mathbf{u}_{\textup{eq}}+ \delta \bar{\mathbf{u}}^M, \textup{diag}(R_{\textup{eq}}+\delta R^M))$ can be written as,
\begin{eqnarray}
\textup{KL}(\pi_\delta,\pi_\delta^M) =  \textup{KL}(\pi_{\textup{eq}},\pi^M_{\textup{eq}}) + \frac{1}{2} \sum_{k\in \cI} R_{\textup{eq},k}^{-1}(\delta \bar{\mathbf{u}}_k - \delta \bar{\mathbf{u}}^M_k)^2+ \frac{1}{4} \sum_{k\in \cI} R_{\textup{eq},k}^{-2} (\delta R_k - \delta R^M_k)^2 + O(\delta^3),\label{KLdiv}
\end{eqnarray} 
where $\delta \bar{\mathbf{u}}_k$ and $\delta\bar{\mathbf{u}}_k^M$ denote the $k$-th component of the mean responses $\delta\bar{\mathbf{u}}$ and $\delta \bar{\mathbf{u}}^M$, respectively, and $R_{\textup{eq},k}, \delta R_k$, and $\delta R_k^M$ denote the $k$-th diagonal component of the matrices $R_{\textup{eq}}, \delta R$, and $\delta R^M$, respectively.
{The choice of fitting only the diagonal entries of the response statistics is reasonable when the covariance is diagonally dominant with small non-diagonal entries. In practice, an accurate proxy of the non-diagonal entries of $R_{\textup{eq}}$ may not be available since such accurate training data may require a simulation with a very large ensemble size especially if the full model is high-dimensional. So, fitting to inaccurate non-diagonal entries in $R_{\textup{eq}}$ may introduce additional errors. In Section~\ref{sec4.4}, we show that fitting to only the diagonal entries of $R_{\textup{eq}}$ with the following loss function still produces an accurate estimation for the variance response.}

Since $\textup{KL}(\pi_{\textup{eq}},\pi^M_{\textup{eq}})=0$ by equilibrium consistency, we propose the following loss function,
\begin{equation}
\textup{L}(\theta) = \sum_{j=1}^T\left( \frac{1}{2} \sum_{k\in \cI} R_{\textup{eq},k}^{-1}(\delta \bar{\mathbf{u}}_k(t_j) - \delta \bar{\mathbf{u}}^M_k(t_j;\theta))^2+ \frac{1}{4} \sum_{k\in \cI} R_{\textup{eq},k}^{-2} (\delta R_k(t_j) - \delta R^M_k(t_j,\theta))^2\right),\label{empirical_loss}
\end{equation}  
which measures the signal and dispersion contribution at discrete time indices $\{t_j =j\Delta t\}_{j=0,\ldots, T}$ to the discrepancy of the response mean and variance statistics between the underlying dynamics and the reduced-order model to be fitted. We specify the loss function to depend on $\theta$, denoting parameters in the class of models used to approximate, $\mathbf{\Theta}^m$ and $Q^M$. When a neural-network model is used, then $\theta$ corresponds to the neural-network model parameters. 

From the classical supervised learning perspective { \cite{mohri2018foundations}, the loss function \eqref{empirical_loss} for the regression problem compares the real-valued labels},
\begin{equation}
    y:=\left(\delta \bar{\mathbf{u}}_k(t_j),\delta R_k(t_j)\,:\, \forall j \in \{1,\ldots, T \},k\in\cI\right),\label{label}
\end{equation}
the statistical responses of the underlying dynamics in \eqref{eq:abs_formu}, to the predicted labels, 
\begin{equation}
y^M:= \left(\delta \bar{\mathbf{u}}_k^M(t_j),\delta R_k^M(t_j)\,:\, \forall j \in \{1,\ldots, T \},k\in\cI\right)\label{predictedlabel}
\end{equation}
the statistical response induced by the reduced-order model in \eqref{generalclosuremodel}. For convenience of the discussion, let us define the operator $\mathcal{M}$ as $y^M = \mathcal{M}(x)$, where $x$ denotes the initial conditions of \eqref{generalclosuremodel} (which will include the appropriate inputs for $\mathbf{\Theta}^m$ and $Q^M$). We will specify the input variable $x$ in Section~\ref{sec3.4} corresponding to a specific reduced-order model accounting for the inputs of $\mathbf{\Theta}^m$ and $Q^M$. With this notation, we
write the loss function $L(\theta):=L(\theta,y,y^M)$ to emphasize its dependence on the label \eqref{label} and predicted label  \eqref{predictedlabel}. The supervised machine learning training corresponds to minimizing the following empirical risk function,
\begin{equation}
R_n(\theta) := \frac{1}{n}\sum_{i=1}^n \textup{L}(\theta,y_i,y^M_i),\label{empirical_risk}
\end{equation}
which is an empirical average of the loss function over $n$ training data $(x_i,y_i)_{i=1,\ldots, n}$. Here, we should emphasize that $y_i^M = \mathcal{M}(x_i)$ is the predicted {response statistics (real-valued label)} corresponds to the input $x_i$. In Section~\ref{sec3.4}, we will specify the input variable $x$ of a specific example of \eqref{generalclosuremodel} and provide a pseudo-algorithm to evaluate the operator $\mathcal{M}$. In Section~\ref{subsec:training_data}, we will provide more detailed discussion on the generation of training data.

\section{Machine learning strategies for modeling unresolved structures with strong instability\label{sec:Machine-learning-strategies}}

To illustrate the key idea in the data-driven modeling framework to
capture leading statistics, we display the detailed construction of the 
general model described in Section \ref{sec:General-mathematical-formulation} in a step-by-step fashion on the L-96 system as one representative example. First, we start with a simpler case only including homogeneous statistics. Then, the inhomogeneous model is developed by adding additional structures subject to the inhomogeneous damping and forcing effects. Especially in modeling systems with { chaotic} dynamics, a crucial issue is to construct { stable approximate dynamical equations that can avoid} the inherent instability in the system.

\subsection{Lorenz '96 system as a representative test model}

In this section, we realize the closure modeling approach described in Section~\ref{sec:General-mathematical-formulation} on a simple prototypical example that exhibits a range of statistical features that arose in {chaotic dynamics.}  Particularly, we consider the 40-dimensional Lorenz '96 (L-96) system \cite{lorenz:96} of state variables $\mathbf{u}=\left(u_{1},u_{2},...,u_{N}\right)^\top$
with general {spatially} inhomogeneous damping and forcing, 
\begin{equation}
\frac{du_{j}}{dt}=\left(u_{j+1}-u_{j-2}\right)u_{j-1}-d_{j}\left(t\right)u_{j}+f_{j}\left(t\right),\;j=1,\cdots,N=40.\label{eq:L96} 
\end{equation}
This ODE system is defined with a periodic boundary condition mimicking geophysical
{ weather dynamics on a midlatitude belt of roughly 32,000 km. The choice of $N=40$ grid points corresponds to a spatial discretization of about 800 km which is the length scale of Rossby radius observed in nature.} Various statistical features that reflect the real observations in nature can be generated by the simple model (\ref{eq:L96}). Especially, inhomogeneous processes
are introduced by the spatially varying damping and forcing effects
$d_{j}$ and $f_{j}$ {as a generalization to the standard L-96 model configuration with uniform damping $d_j\equiv d_{\mathrm{eq}}=1$ and forcing $f_j\equiv F_{\mathrm{eq}}=8$ (some explicit forms of the inhomogeneous forcing and damping that we use in our numerical are illustrated in Fig.~\ref{fig:Different-types} in Section~\ref{sec:Predicting-leading-order})}. This will lead to more complicated inhomogeneous
statistics in the mean modes as well as the non-zero off-diagonal
covariances. To compare with the abstract form (\ref{eq:abs_formu}), we can write the linear and quadratic operators for L-96 system
as
\[
\mathcal{L}+\mathcal{D}=\mathrm{diag}\left(-d_{1},\cdots,-d_{N}\right),\quad B\left(\mathbf{u},\mathbf{v}\right)=\left\{ u_{i-1}^{*}\left(v_{i+1}-v_{i-2}\right)\right\} _{i=1}^{N}
\]
and project the state variables onto the Fourier basis $\mathbf{e}_{k}=\left\{ e^{i2\pi k\frac{l}{N}}\right\} _{l=1}^{N}$
considering the periodic boundary condition.

We aim to deduce moment closure equations for (\ref{eq:L96})  that include inhomogeneous
structures in the statistical mean and stochastic fluctuation modes. In order to achieve
this, we project the general inhomogeneous forcing and damping as well as the model state onto each spectral mode such that 
\begin{equation}
\begin{aligned}f_{j}=\: & \hat{f}_{0}+\sum_{k\ne0}\hat{f}_{k}e^{i2\pi k\frac{j}{N}},\quad\quad d_{j}=\hat{d}_{0}+\sum_{k\ne0}\hat{d}_{k}e^{i2\pi k\frac{j}{N}},\\
u_{j}(t,\omega)=\: & \bar{u}_{j}\left(t\right)+\sum_{|k|\leq\nicefrac{N}{2}}Z_{k}\left(t;\omega\right)e^{i2\pi k\frac{j}{N}}.
\end{aligned}
\label{eq:mode_spec}
\end{equation}

Above, we denote the homogeneous components of the forcing and damping as $\hat{f}_{0}$ and $\hat{d}_{0}$, respectively, corresponding to the Fourier mode $k=0$. Notice that 
in the decomposition in (\ref{eq:mode_spec}), the state variable
$u_{j}=\bar{u}_{j}+u_{j}^{\prime}$ is 
decomposed into the statistical
mean $\bar{u}_{j}$ and the fluctuations $u_{j}^{\prime}$, which is then written as a linear combination of the 
fluctuation modes $Z_{k}$ in Fourier coordinates.  We will further decompose the mean state into the contributions of the homogeneous and inhomogeneous terms as,
\[
\bar{u}_{j}(t)=\hat{u}_{0}(t)+\sum_{|k|\leq\nicefrac{N}{2}}\hat{u}_{k}\left(t\right)e^{i2\pi k\frac{j}{N}},
\]
where $\hat{u}_k(t)$ corresponds to the $k$-th Fourier mode of the mean $\bar{u}_j$. If $\bar{u}_j = \bar{u}$ is spatially homogeneous, then the zeroth mode $\hat{u}_{0}(t)$ is precisely the homogeneous mean $\bar{u}(t)$. This observation implies that the non-zero Fourier modes characterize the inhomogeneity of the dynamical processes.

\subsubsection{Statistical mean dynamics}

Projecting Equation (\ref{eq:L96}) to different spectral modes,
we obtain the statistical mean equation for the homogeneous and inhomogeneous
components\addtocounter{equation}{0}\begin{subequations}\label{eq:dyn_mean_inhomo}
\begin{align}
\frac{d\hat{u}_{0}}{dt}= &-\bar{d}\hat{u}_{0} -  \sum_{k\neq0}d_{0,k}\hat{u}_{k}+\hat{f}_{0} +\sum_{|k|\leq\nicefrac{N}{2}}\left(\left|\hat{u}_{k}\right|^{2}+\left\langle |Z_{k}|^{2}\right\rangle \right)\gamma_{k},\label{eq:dyn_mean_inhomo1}\\
\frac{d\hat{u}_{k}}{dt}= &- \sum_{|m|\leq\nicefrac{N}{2}}d_{k,m}\hat{u}_{m}+\hat{f}_{k} 
+  \sum_{|m|\leq\nicefrac{N}{2}}\left(\hat{u}_{m}\hat{u}_{k-m}+\left\langle Z_{m}Z_{k-m}\right\rangle \right)\gamma_{m}^{*}e^{-i2\pi\frac{k}{N}},\label{eq:dyn_mean_inhomo2}
\end{align}
\end{subequations}with the uniform damping rate $\bar{d}=\frac{1}{N}\sum_{j}d_{j}$,
and the damping coefficients for each inhomogeneous mode $d_{k,m}=\frac{1}{N}\sum_{j}d_{j}e^{i2\pi\left(m-k\right)\frac{j}{N}}=\hat{d}_{k-m}$.
The nonlinear coupling between different scales is connected by the
coefficient $\gamma_{k}=e^{-i\frac{4\pi k}{N}}-e^{i\frac{2\pi k}{N}}$.
Notice that the first equation (\ref{eq:dyn_mean_inhomo1}) only contains
homogeneous dynamics (no cross-correlation between different wavenumbers
$k$). In addition to the homogeneous mean mode $\hat{u}_0$,
we also need to compute the inhomogeneous mean modes $\hat{u}_{k}$
if inhomogeneous forcing and damping effects are included.

\subsubsection{Stochastic coefficient dynamics}

The dynamical equation for the stochastic coefficients can be attained by subtracting the mean dynamics (\ref{eq:dyn_mean_inhomo})
from the original equation (\ref{eq:L96}) and subsequently projecting it to each spectral mode. Following these steps, we have the governing equation for the stochastic coefficients $Z_{k}$ as,
\begin{equation}
\begin{aligned}\frac{dZ_{k}}{dt}=  -\sum_{|m|\leq\nicefrac{N}{2}}d_{k,m}Z_{m}+\sum_{|m|\leq\nicefrac{N}{2}}\mu_{k,m}\hat{u}_{k-m}Z_{m}
  +\sum_{|m|\leq\nicefrac{N}{2}}\left(Z_{m}Z_{k-m}-\left\langle Z_{m}Z_{k-m}\right\rangle \right)\gamma_{m}^{*}e^{i2\pi\frac{-k}{N}},
\end{aligned}
\label{eq:dyn_coeff_inhomo}
\end{equation}
with the coupling coefficient $\mu_{k,m}=e^{i2\pi\frac{k-2m}{N}}+e^{i2\pi\frac{2m-k}{N}}-e^{i2\pi\frac{m-2k}{N}}-e^{i2\pi\frac{-k-m}{N}}$.
On the right-hand-side of (\ref{eq:dyn_coeff_inhomo}), the first term denotes linear damping, the second term characterizes the coupling through the homogeneous and inhomogeneous means, and the third term characterizes the nonlinear
coupling between the fluctuation modes between different scales.

For a complete investigation of the energy transferring mechanism
subject to linear and nonlinear interactions, we can also derive the
corresponding dynamical equation for the covariance $R_{kl}=\left\langle Z_{k}Z_{l}^{*}\right\rangle $
according to (\ref{eq:cov_dyn})
\begin{equation}
\begin{aligned}\frac{dR_{kl}}{dt}= & -2\bar{d}R_{km}-\left(\gamma_{-k}+\gamma_{-k}^{*}\right)\hat{u}_{0}R_{kl}\\
 & -\sum_{m\neq k}\left(d_{k,m}R_{ml}+d_{l,m}^{*}R_{km}\right)+\sum_{m\neq k}\left(\mu_{k,m}\hat{u}_{k-m}R_{ml}+\mu_{l,m}^{*}\hat{u}_{l-m}^{*}R_{km}\right)\\
 & +\sum_{m\ne0}\left\langle Z_{m}Z_{k-m}Z_{l}^{*}\right\rangle \gamma_{m}^{*}e^{i2\pi\frac{-k}{N}}+\left\langle Z_{m}^{*}Z_{l-m}^{*}Z_{k}\right\rangle \gamma_{m}e^{i2\pi\frac{l}{N}}.
\end{aligned}
\label{eq:dyn_cov_l96}
\end{equation}
The homogeneous effects due to damping and mean interaction
are summarized in the first row of (\ref{eq:dyn_cov_l96}). The inhomogeneous
damping and mean interactions are shown in the second row of (\ref{eq:dyn_cov_l96}).
Higher-order feedbacks from the third-order moments with non-Gaussian statistics
among all the spectral modes enter the covariance equation in the
third row of (\ref{eq:dyn_cov_l96}).

In the following, we describe the step-by-step construction of the
data-driven reduced-order model on the L-96 system as a canonical
example, following the same reduced modeling approach that is stated for the more abstract model in (\ref{eq:mean_dyn}) and (\ref{eq:coeff_dyn})
on the particular case (\ref{eq:dyn_mean_inhomo})
and (\ref{eq:dyn_coeff_inhomo}).

\subsection{Hybrid statistical-stochastic model for homogeneous statistics}

\begin{figure}
\includegraphics[scale=0.37]{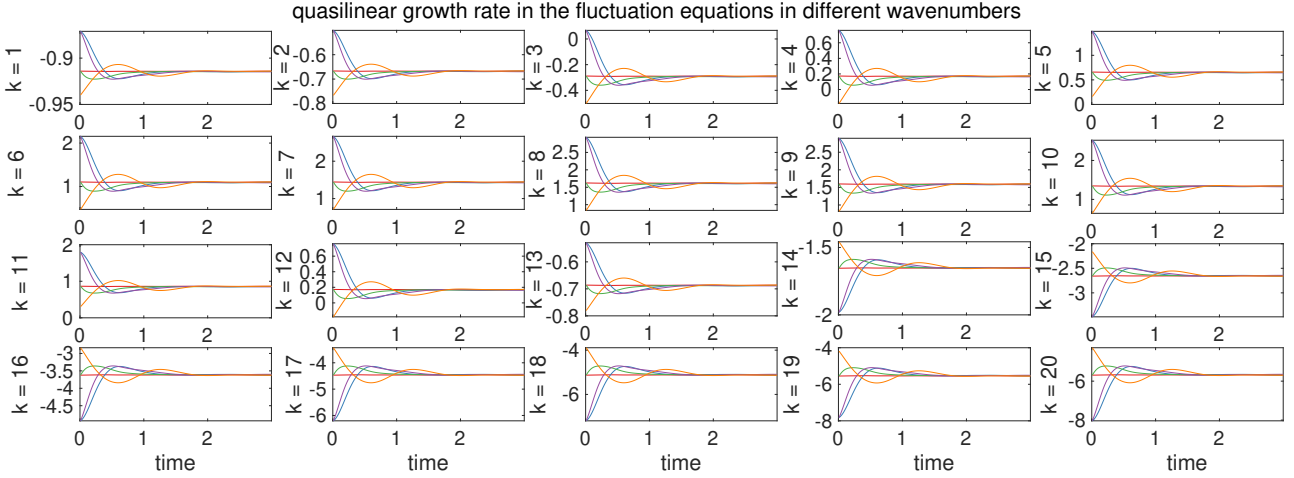}
\caption{The time evolution of the quasilinear growth rate computed for each spectral
mode $k$ of the L-96 model. Different lines are subject to the initial
perturbations described in Section \ref{subsec:training_data}. }
\label{fig:instability}
\end{figure}

We start with the simple model set up with homogeneous damping and
forcing, $d_{j}:= \gamma,f_{j}:= f$, in (\ref{eq:mode_spec})
together with homogeneous initial perturbations $u_{0,j}:= u_{0}$. 
In this case, the mean and fluctuation equations in (\ref{eq:dyn_mean_inhomo})
and (\ref{eq:dyn_coeff_inhomo}) can be simplified as \addtocounter{equation}{0}\begin{subequations}\label{eq:model_homo}
\begin{align}
\frac{d\bar{u}}{dt} & =-\gamma\bar{u}+\sum_{k} R_k \gamma_{k}+f,\label{eq:model_homo1}\\
\frac{dZ_{k}}{dt} & =-\left(\gamma+\gamma_{k}\bar{u}\right)Z_{k}+\sum_{m\neq 0}Z_{m}Z_{k-m}\gamma_{m}^{*}e^{i2\pi\frac{-k}{N}},\label{eq:model_homo2}
\end{align}
\end{subequations}with $\gamma_{k}=e^{-i\frac{4\pi k}{N}}-e^{i\frac{2\pi k}{N}}$,  $\bar{u}= \hat{u}_0$, $\bar{d}=\gamma$, 
and $R_{k}:=R_{kk}=\left\langle Z_{k}Z_{k}^{*}\right\rangle $ denotes the variance of the stochastic coefficient $\hat{Z}_{k}\left(t;\omega\right)$. 
Under the homogeneous statistics, the statistical mean state becomes
a scalar and the covariance matrix becomes diagonal, that is,
\[
\bar{u}_{j}=\bar{u} =  \hat{u}_0,\;\hat{u}_{k}:=0, k\neq0,\;\mathrm{and}\; R_{kl}=R_{k}\delta_{kl}.
\]
Thus we do not need to consider the inhomogeneous mean equation (\ref{eq:dyn_mean_inhomo2})
involving $\hat{u}_{k}$ and the cross-correlations between different
spectral mode $\left\langle Z_{k}Z_{l}^{*}\right\rangle ,k\ne l$.
On the other hand, nonlinear dynamics and non-Gaussian statistics
still play a central role due to the strongly coupled feedbacks in
equations (\ref{eq:model_homo}). Different scales are mixed in the
feedbacks with summations over all the wavenumbers. In particular,
the system may contain strong internal instability through the mean-fluctuation
interactions. For example, in (\ref{eq:dyn_mean_inhomo2}) strong
positive growth rate will occur when $\hat{u}_0 = \bar{u}>0$ for modes with
$\mathfrak{Re}\gamma_{k}<0$. To illustrate this, we plot in Figure \ref{fig:instability}  the quasilinear growth rate $-\left(\gamma+\gamma_{k}\bar{u}\right)$ of each spectral mode in the L-96 model, subject to different initial perturbations (that we will describe in Section~\ref{subsec:training_data}). Positive value implies instability of the mode. We notice that the instabilities occur on a wide range of modes depending on the {initial value perturbations and their intermittent occurences} create a practical challenge for learning a {dynamically stable model and accurate prediction of model statistics}. 

\subsubsection{Direct modeling of the unresolved feedbacks}

Using the hybrid statistical-stochastic model (\ref{eq:model_homo}),
it requires the computation of the statistical expectation $R_{k}=\left\langle Z_{k}Z_{k}^{*}\right\rangle $
for the mean equation (\ref{eq:model_homo1}) from the solution of
the stochastic equation (\ref{eq:model_homo2}). In practice, this
can be achieved by ensemble simulation of the stochastic coefficients
$Z_{i}$.  {With the ensemble approximation in \eqref{R^M_empirical}, the analog of the closure model in \eqref{generalclosuremodel} in this example for the homogeneous case is given by:}
\begin{equation}
\begin{aligned}\frac{d\bar{u}^M}{dt} & =-\gamma\bar{u}^M+\sum_{k\in\cI }\left(\frac{1}{M-1}\sum_{i=1}^{M}Z_{k}^{M,\left(i\right)}(Z_{k}^{M,\left(i\right)})^*\right)\gamma_{k}+\Theta^{m}+f,\\
\frac{dZ_{k}^{M,\left(i\right)}}{dt} & =-\left(\gamma+\gamma_{k}\bar{u}^M\right)Z_{k}^{M,\left(i\right)}+\Theta_{k}^{v},\quad k\in\cI,\;i=1,\cdots,M.
\end{aligned}
\label{eq:model_ml_red1}
\end{equation}
In (\ref{eq:model_ml_red1}), $\mathbf{Z}^{M,\left(i\right)}= (Z_k^{M,(i)}),k\in\cI, i=1,\cdots,M$
is each (independent) ensemble member in the ensemble simulation of
the fluctuation equations, {$\cI$ is the set containing the resolved fluctuation modes corresponding to the largest variances}. Its statistical feedback in the mean equation
is approximated by the empirical ensemble average
in \eqref{R^M_empirical}. 

In the above model, we consider $\cI$ to include modes {with largest variances but still only accounting for a small portion of the total energy} (that is, $k=7-13$, see Figure~\ref{fig:Equilibrium-stat}(a))   so that the  leading order dynamics can be replicated. The resolved fluctuations provide the explicit variance feedback and nonlinear coupling in the
mean and fluctuation equations, whereas the data-driven component behaves as a higher-order correction. Comparing the reduced-order
model (\ref{eq:model_ml_red1}) with the full exact model (\ref{eq:model_homo}),
the process $\Theta^{m}$ models the variance feedback to the mean
dynamics among all the unresolved small-scale modes $k \in \{-N/2+1,\ldots, N/2\}\backslash \cI$.
In addition, $\Theta^{m}$ also accounts for the approximation error
in the resolved variances from the empirical ensemble average from
 a finite sample size, $M$. In the stochastic equations
for $Z_{k}^M$, the model $\Theta_{k}^{v}$ approximates the  total
contribution from the nonlinear coupling among all the fluctuation
modes.

\subsubsection{Stable {dynamical equations} with effective damping and noise}

First, to overcome the instabilities occurring on unstable modes, where $\gamma+\gamma_{k}\bar{u}<0$ (see Fig. \ref{fig:instability}), we follow the strategy discussed in Section~\ref{sec2.2.1}.
Next, {instead of an extensive pointwise calibration of each stochastic
trajectory of $Z_{k}^{\left(i\right)}$, we propose to only measure
the error in the ensemble statistics discussed in Section~\ref{sec2.2.2}, so that the high computational cost in training is effectively avoided while the statistical accuracy is also maintained.} 
Notice that for
the homogeneous case, the dynamical equation for the covariance matrix $R$ in \eqref{eq:cov_dyn} can be simplified. Particularly, $R$ only has nontrivial diagonal components $R_k$ that satisfied,
\begin{eqnarray}
\frac{dR_{k}}{dt}=-2\gamma R_{k}-\left(\gamma_{k}+\gamma_{k}^{*}\right)\bar{u}R_{k}+Q_{F,k},\label{variance_homo_dyn}
\end{eqnarray}
where $Q_{F}$ is a diagonal matrix for the high-order statistical
nonlinear fluxes.

{In this homogeneous case, the proposed statistical-stochastic model in \eqref{generalclosuremodel} is simplified to,}
\begin{equation}
\begin{aligned}
\frac{d\bar{u}^M}{dt} & =-\gamma\bar{u}^M+\sum_{k\in\cI }\left(\frac{1}{M-1}\sum_{i=1}^{M}Z_{k}^{M,\left(i\right)}(Z_{k}^{M,\left(i\right)})^*\right)\gamma_{k}+\Theta^{m}+f,\\
\frac{dZ_{k}^{M,\left(i\right)}}{dt} & =-\left(\gamma+\gamma_{k}\bar{u}^M\right)Z_{k}^{M,\left(i\right)}-D_k^M Z_{k}^{M,\left(i\right)} + \Sigma_{k}^M\dot{W}_{k}^{\left(i\right)} ,\quad k\in\cI,\;i=1,\cdots,M,
\end{aligned}
\label{eq:model_ml_red2}
\end{equation}
where $D_k$ and $\Sigma_k$ are parameterized as in \eqref{parameterizations}. 
\begin{equation}
\begin{aligned}
D_{k}^{M} & =-\frac{\min\left\{ Q_{k}^{M},0\right\}}{{2}R_{\mathrm{eq},k}},\\
\Sigma_{k}^{M} & =\sqrt{\max\left\{ Q_{k}^{M},0\right\} }.
\end{aligned}
\label{eq:separation}
\end{equation}
We should point out that the component-wise decomposition \eqref{eq:separation} is only possible since the statistics is homogeneous, and thus, avoiding an expensive matrix decomposition to identify the positive and negative definite components, {$(Q_M)^+$} and {$(Q_M)^-$,} respectively, that satisfy {$Q_M:=(Q_M)^+-(Q_M)^-$} for non-diagonal matrix {$Q_M$.}

In section~\ref{sec3.4}, we will specify the class of machine learning models to identify $\Theta^{m}_k$ and $Q_{k}^M$ in terms of time delay embedding of these variables, respectively, in addition to the time delay embedding of the mean and variance. Before discussing this, we consider a slight modification to the closure model above to accommodate for inhomogeneous statistics in the next section.

\subsection{The statistical-stochastic model for inhomogeneous statistics}

\begin{figure}
\subfloat[equilibrium mean \& variance spectrum]{\includegraphics[scale=0.38]{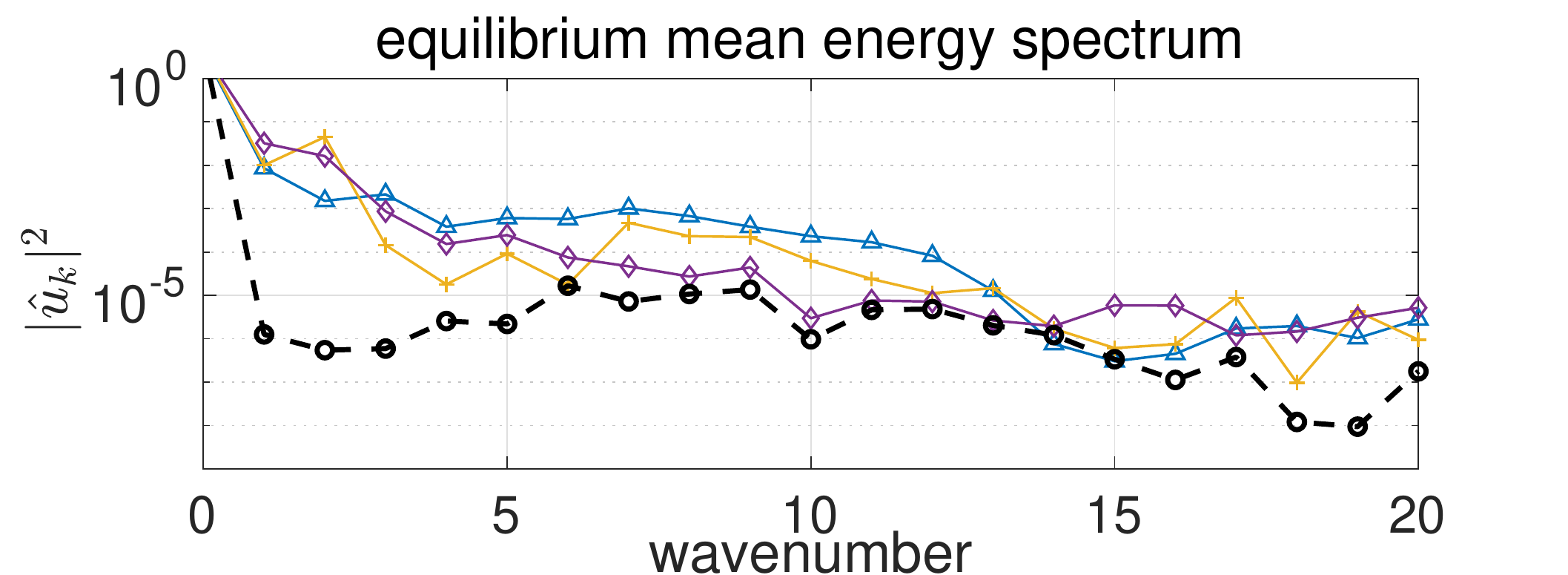}\includegraphics[scale=0.38]{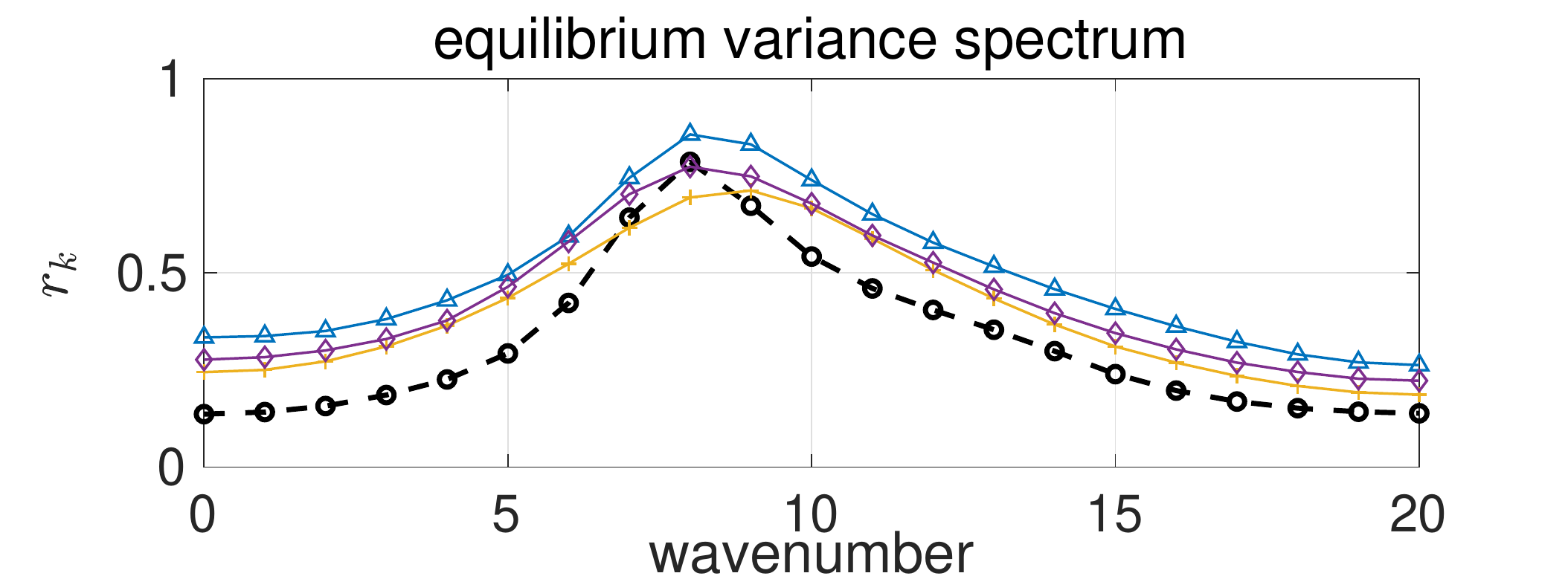}

}

\subfloat[equilibrium covariance matrix]{\includegraphics[scale=0.38]{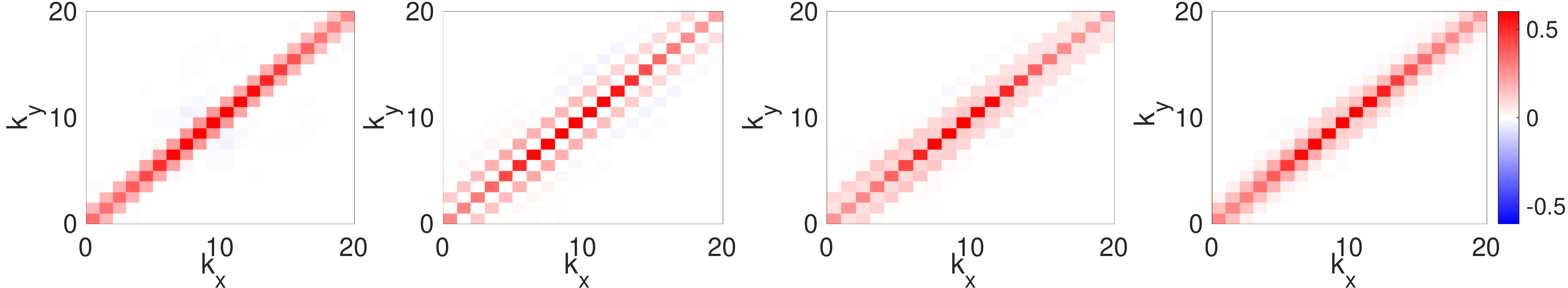}

}

\caption{Equilibrium statistics of the L-96 model with inhomogeneous perturbations.
First row: equilibrium spectra of energy in the mean and variance
covariance (unperturbed homogeneous case in dashed line). Second row:
the equilibrium covariance under several inhomogeneous forcing and
damping effects.\label{fig:Equilibrium-stat}}
\end{figure}

Next, we consider to predict the mean and variance responses under a more general case { with inhomogeneous statistics introduced by spatially inhomogeneous forcing
and initial perturbations}. For this case, we have the additional observations
for the inhomogeneous equations (\ref{eq:dyn_mean_inhomo}) and (\ref{eq:dyn_coeff_inhomo}):
\begin{itemize}
\item The homogeneous mean state $\bar{u}=\hat{u}_{0}$ is subject to feedbacks
from not only the variances $R_{k}$ (as in the homogeneous
case), but also the energy in the inhomogeneous mean states $\left|\hat{u}_{k}\right|^{2}$;
\item The inhomogeneous mean modes $\hat{u}_{k}$ are subject to the cross
interactions between the mean states $\hat{u}_{0}\hat{u}_{k}$ and
the cross-covariances $\left\langle Z_{k}Z_{0}\right\rangle $;
\item The stochastic coefficients $Z_{k}$ are subject to the cross interactions
between the mean and the fluctuation modes as well as the nonlinear
coupling between different wavenumber modes.
\end{itemize}
In the inhomogeneous case, it is expensive to resolve the cross
interaction terms between the entire spectrum. In Figure \ref{fig:Equilibrium-stat},
we plot the typical spectra for the mean and variance under several
different inhomogeneous perturbations. First, we should point out that the homogeneous
mean $\hat{u}_{0}$ and variance $R_{k}$ for $7\leq k 
\leq 13$ are still dominant under various inhomogeneous forces.  While including these non-trivial modes in $\cI$ is sufficient for homogeneous modeling, excluding other modes (such as $1\leq k 
\leq 6$) whose mean energy spectra are significantly increased under inhomogeneous forces will produce a poor statistical recovery. In general, the reduced model should include modes that are significantly excited by the inhomogeneous perturbations, which makes the modeling choice slightly more complicated than that of the homogeneous case. From Figure~\ref{fig:Equilibrium-stat}(b), we also notice that the covariance matrices of the perturbed dynamics are diagonally banded with the detailed structure depending crucially on the perturbations. While the non-diagonal components are non-negligible, they are much smaller compared to the diagonal components. This scale separation poses an additional computational challenge for an accurate estimation of the non-diagonal covariance components, which is crucial for stable modeling of the inhomogeneous components as shown in \eqref{eq:dyn_mean_inhomo2}.

Given these statistical features, we consider a diagonal closure model {for the feedback from the higher-order moments}, measuring only the statistics in the inhomogeneous
mean and diagonal variances. Particularly,
we consider the dynamical closure equations for the homogeneous mean
$\hat{u}_{0}$, inhomogeneous mean $\hat{u}_{k}$, and the fluctuation
modes $Z_{k}$ corresponding to the resolved subset $k\in \cI$ as follows:
\begin{equation}
\begin{aligned}\frac{d\hat{u}_{0}^M}{dt}= & -\sum_{j\in\cI}d_{0,j}\hat{u}_{j}^M+\bar{f}+\sum_{k\in\cI}\left(\left|\hat{u}_{k}^M\right|^{2}+R_{k}^{M}\right)\gamma_{k}\;+\Theta_{0}^{m},\\
\frac{d\hat{u}_{k}^M}{dt}= & -\sum_{\ell\in\cI}d_{k,\ell}\hat{u}_{\ell}^M+\hat{f}_{k}-\gamma_{k}\hat{u}_{0}^M\hat{u}_{k}^M\;+\Theta_{k}^{m},\\
\frac{dZ_{k}^{M,\left(i\right)}}{dt}= & -\sum_{\ell\in\cI}d_{k,\ell}Z_{\ell}^{M,\left(i\right)}-\gamma_{k}\hat{u}_{0}^M Z_{k}^{M,\left(i\right)}-D_{k}^{M}Z_{k}^{M,\left(i\right)}+\Sigma_{k}^{M}\dot{W}_{k}^{\left(i\right)},
\end{aligned}
\label{eq:model_ml_red3}
\end{equation}
which is an example of \eqref{generalclosuremodel}. Here $d_{k,m}$ are defined as in \eqref{eq:dyn_mean_inhomo1}-\eqref{eq:dyn_mean_inhomo2}. 
The variance feedback $R_k^M$ in the mean equation is defined as the diagonal component of \eqref{R^M_empirical} attained with $M$ samples. Also, the feedbacks from the non-diagonal
covariance entries are not computed explicitly in the second equation in \eqref{eq:model_ml_red3} due to their relatively
small amplitudes. Following the homogeneous case, we introduce $\Theta_{0}^{m},\Theta_{k}^{m}$ to
account for the truncation error and the feedbacks in the homogeneous and inhomogeneous mean state
from unresolved small-scale processes. The stabilizing decomposition
for effective damping $D_k^M$ and noise $\Sigma_k^M$ is constructed exactly as in
(\ref{eq:separation}),
except that the model $Q_k^M$ in \eqref{eq:separation} is fitted to the higher-order statistical flux $Q_{F,k}$ induced by the inhomogeneous dynamics for the variance component,
\begin{equation}
\frac{dR_{k}}{dt}=-\sum_{m\in\cI}\left(d_{k,m}R_{mk}+d_{k,m}^{*}R_{km}\right)-\left(\gamma_{k}+\gamma_{k}^{*}\right)\bar{u}R_{k}\;+Q_{F,k},\label{eq:model_var_diag}
\end{equation}
replacing \eqref{variance_homo_dyn} of the homogeneous case.
In this diagonal closure modeling, we should point out that
the element-wise decomposition (\ref{eq:separation}) can still be performed and thus avoiding
the matrix decomposition {$Q_M:=(Q_M)^+-(Q_M)^-$} for non-diagonal case.

In this inhomogeneous model, we make a final remark that the proposed closure $Q^{M}_k$ is to account for modeling error induced by: i) the higher-order moment feedback to the variance; ii) the variances of the unresolved modes $k \in \cI^{c}$; and iii) the neglected cross-covariances $R_{k\ell},k\ne \ell$ that are not explicitly computed in the model. In Section~\ref{sec:Predicting-leading-order}, we will empirically show that the proposed diagonal reduced-order model, which is numerically efficient, does not introduce significant error to the prediction of the mean and variance statistics.

{
\noindent {\bf An extension from the previous work.}
In the previous work where only modeling homogeneous dynamics is considered \cite{qi2022machine}, in addition to not having non-homogeneous mean modes, $\{\hat{u}_k\}_{k\in \cI, k\neq 0}$, we also employed a closure to the diagonal model for $R_k$ in \eqref{variance_homo_dyn}. In that work, we employed the same fitting as in \eqref{eq:separation} to $Q_{F,k} \approx Q_k^M$, and considered the following closure model for the variance dynamics,
\begin{eqnarray}
\frac{dR_{k}^M}{dt} = -2\gamma R_{k}^M-\left(\gamma_{k}+\gamma_{k}^{*}\right)\bar{u}R_{k}^M- 2D_k^M R_k^M + (\Sigma_k^M)^2.\label{eq:model_var_diag2}
\end{eqnarray}
Compare to the dynamics of $Z_{k}^{M,(i)}$ in \eqref{eq:model_ml_red3}, the diagonal model in \eqref{eq:model_var_diag2} does not capture the feedback from different modes through $\sum_{\ell\in\cI}d_{k,\ell}Z_{\ell}^{M,\left(i\right)}$ in \eqref{eq:model_ml_red3}. The current approach with \eqref{eq:model_ml_red3} is analogous to extending the previous approach in \eqref{eq:model_var_diag2}
to include the non-diagonal second-order statistical interactions. Indeed, one can see that in the limit of large ensemble size, the covariance dynamics of \eqref{eq:model_ml_red3} is identical to that achieved by imposing the diagonal closure in \eqref{eq:separation} to approximate the third-order moments in \eqref{eq:model_var_diag}, 
\begin{equation}
\begin{aligned}
\frac{dR_{kl}}{dt} = & -2\bar{d}R_{km}-\left(\gamma_{-k}+\gamma_{-k}^{*}\right)\hat{u}_{0}R_{kl}\\
 & -\sum_{m\neq k}\left(d_{k,m}R_{ml}+d_{l,m}^{*}R_{km}\right)+\sum_{m\neq k}\left(\mu_{k,m}\hat{u}_{k-m}R_{ml}+\mu_{l,m}^{*}\hat{u}_{l-m}^{*}R_{km}\right) \\
 & + (- 2D_k^M R_k^M + (\Sigma_k^M)^2) \delta_{kl}.\label{limitingcov}
 \end{aligned}
\end{equation}
While one can use this closure model for $R_{kl}$, we choose to consider the fluctuation dynamics for $Z_k^M$ for the following reasons. First, the closure model for the fluctuation components allows one to estimate statistics other than covariance $R^\cI$, whenever improved parameterization of the unresolved feedback from higher-order moments becomes available. Second, using the ensemble approach, the empirical estimate of the covariance is always symmetric and positive. While this may not pose any serious issue with the current diagonal closure in \eqref{limitingcov} provided the $K\times K$ covariance dynamics is numerically integrated with an adequate ODE solver and time step, preserving the symmetry can be challenging when more non-diagonal parameterization model $Q_M$ becomes available. Third, we note that the fluctuation dynamics in \eqref{eq:model_ml_red3} consists of an ensemble of independent $K-$dimensional model for $\{Z_k^{M,(i)}\}_{k\in \cI}$. This independent structure allows one to employ a parallel computation for each ensemble member in the model integration to reduce the time complexity when $M>K$, and possibly be faster than directly integrating the fully coupled $K\times K$ system in \eqref{limitingcov}. Particularly, parallel computing will be natural when the neural-network model for $Q_{k}^M$ is employed on a GPU cluster.

}

\subsection{Neural network model for the unresolved processes}\label{sec3.4}

To parameterize $\left\{ \Theta^{m},Q_{k}^{M}\right\} $ in \eqref{eq:model_ml_red2} or $\left\{ \Theta_{0}^{m},\Theta_{k}^{m},Q_{k}^{M}\right\} $ in \eqref{eq:model_ml_red3}, we consider using a non-Markovian closure model to learn these terms, following our previous works in modeling variance closure \cite{qi2022machine} and trajectory of partially observed discrete-time ergodic Markov chain \cite{HJLY:19}. While such a general formulation can be theoretically justified in the context of predicting time-evolution of state variables (using the discrete Mori-Zwanzig representation and the time delay Taken's embedding theory \cite{HJLY:19,gilani2021kernel}), the dimension of the theoretically justifiable observable in the current application is too high for a numerically tractable implementation. 
Particularly, for the statistical-stochastic models in \eqref{eq:model_ml_red2} 
accounting $M$ ensemble members, identifying a model of $K=|\cI|$ variables that depends on $L$-time delay corresponds to estimating {a map with $(MK+1)L$ variables}, which is a very high-dimensional estimation problem since $M \gg 1$ is needed for reasonably accurate ensemble estimations. 
{Rather than learning the entire complicated dynamical processes, we identify the input of the non-Markovian map to reflect some explicit expression of $\left\{ \Theta_{0}^{m},\Theta_{k}^{m},Q_{k}^{M}\right\}$ as reported in  \eqref{eq:dyn_mean_inhomo} and design the neural network only to learn these unresolved processes}. Particularly, since these variables are ultimately functions of the statistical quantities, we will identify them as time delay mappings of the mean and variances of the resolved components in addition to the time-delay of the variable of interest, neglecting their dependence on the smaller non-diagonal covariance components.

Following the work in \cite{qi2022machine}, we will consider the class of residual network with {vanilla LSTM architecture \cite{Hochreiter_1997}}. 
For the homogeneous model, the
correction terms in both the mean equation (\ref{eq:model_homo1})
and the fluctuation equations (\ref{eq:model_homo2}) are approximated
by a neural network with residual structure,
\begin{equation}
\begin{aligned}\Theta^{m}\left(t_{\ell+1}\right) & =\Theta^{m}\left(t_{\ell}\right)+\mathrm{LSTM}^{m}\left(\bar{u}\left(t_{\ell-L:\ell}\right),\left\{ R_{k}\left(t_{\ell-L:\ell}\right)\right\} ,\Theta^{m}\left(t_{\ell-L:\ell}\right);\theta\right),\\
Q_{k}^{M}\left(t_{\ell+1}\right) & =Q_{k}^{M}\left(t_{\ell}\right)+\mathrm{LSTM}_{k}^{v}\left(\bar{u}\left(t_{\ell-L:\ell}\right),\left\{ R_{k}\left(t_{\ell-L:\ell}\right)\right\} ,\left\{ Q_{k}^{M}\left(t_{\ell-L:\ell}\right)\right\} ;\theta\right),
\end{aligned}
\label{eq:lstm_homo}
\end{equation}
where we have used the notation $a(t_{\ell-L:\ell}):=\big(a(t_{\ell-L}),a(t_{\ell-L+1}), a(t_\ell))$ for any dependent variable $a$ and $\theta$ to denote the parameters in the LSTM network. Notice that the right-hand-sides of both equations in \eqref{eq:lstm_homo} are $(2K+1)L$ dimensional maps, independent of the ensemble size $M$. For the inhomogeneous case, the unresolved model parameters $\left\{ \Theta_{0}^{m},\Theta_{k}^{m},Q_{k}^{M}\right\} $ can also be learned directly by fitting the LSTM neural networks with analogous residual structure,
that is,
\begin{equation}
\begin{aligned}\Theta_{0}^{m}\left(t_{\ell+1}\right) & =\Theta_{0}^{m}\left(t_{\ell}\right)+\mathrm{LSTM}_{0}^{m}\left[\bar{u}\left(t_{\ell-L:\ell}\right),\left\{ \hat{u}_{k}\left(t_{\ell-L:\ell}\right)\right\} ,\left\{ R_{k}\left(t_{l-m:l}\right)\right\} ,\Theta_{0}^{m}\left(t_{\ell-L:\ell}\right);\theta\right],\\
\Theta_{k}^{m}\left(t_{\ell+1}\right) & =\Theta_{k}^{m}\left(t_{\ell}\right)+\mathrm{LSTM}_{k}^{m}\left[\bar{u}\left(t_{\ell-L:\ell}\right),\left\{ \hat{u}_{k}\left(t_{\ell-L:\ell}\right)\right\} ,\left\{ R_{k}\left(t_{\ell-L:\ell}\right)\right\} ,\left\{ \Theta_{k}^{m}\left(t_{\ell-L:\ell}\right)\right\} ;\theta\right],\\
Q_{k}^{M}\left(t_{\ell+1}\right) & =Q_{k}^{M}\left(t_{\ell}\right)+\mathrm{LSTM}_{k}^{v}\left[\bar{u}\left(t_{\ell-L:\ell}\right),\left\{ \hat{u}_{k}\left(t_{\ell-L:\ell}\right)\right\} ,\left\{ R_{k}\left(t_{\ell-L:\ell}\right)\right\} ,\left\{ Q_{k}^{v}\left(t_{\ell-L:\ell}\right)\right\};\theta \right].
\end{aligned}
\label{eq:lstm_inhomo}
\end{equation}
We should point that since the non-diagonal terms in the covariance are small relative to the diagonal components, we only include the variance components as inputs, and thus, arrive at {a problem of estimating time delay embedding maps with $(3K+1)L$ variables.}

The LSTM model parameters, $\theta$, are attained by minimizing the empirical risk in \eqref{empirical_risk} defined by averaging the loss function $L(\theta,y_i,y_i^M)$ on $n$ training samples of $(x_i,y_i)_{i=1}^n$, where {$x_i$ takes in the sequences of input data $\bar{u}, \hat{u}_{k}, R_k$ and $y_i^M = \mathcal{M}(x_i)$ gives the LSTM output}. In the following pseudo-code, we provide the computational steps for evaluating $y^M = \mathcal{M}(x)$, where the input $x$ corresponding to the statistical-stochastic reduced-order model in \eqref{eq:model_ml_red3} is also stated precisely.
We remark that similar pseudo-code is used for the homogeneous case, where the reduced-order model in \eqref{eq:model_ml_red3} is replaced with \eqref{eq:model_ml_red2} and the LSTM closure models in \eqref{eq:lstm_inhomo} is replaced with \eqref{eq:lstm_homo}. In the homogeneous case, the input $x$ does not have $\{\hat{u}_{k}\}_{k\in\cI}$. 

\begin{algorithm}[ht]
\caption{Evaluating the label $y^M = \mathcal{M}(x,{\theta})$ corresponds to the reduced-order statistical-stochastic model.}\label{alg1}
\begin{algorithmic}
\State {\bf{Input:}} $x$ consists of $\hat{u}_{0}, \hat{u}_{k}, R_k$ at time $t_{-L}, \ldots, t_0$ and $ Z_{k}^{(i)}$ at time $t_{0}$, where $\Delta t = t_{\ell}-t_{\ell-1}$ for all $k\in \cI$ and $i=1,\ldots, M$.  
{($\theta$ denotes the parameters in the LSTM model.)}
\State {\bf{Output:}} $y^M$ consists of $\delta \hat{u}_0^M, \delta \hat{u}_k^M, \delta R_k^M$ for all $k \in \cI$ at times $t_1,\ldots, t_{T}$.
\Require $\ell=1$, $T>0$ 
\While{$\ell < T$}
\begin{itemize}
\item Compute the unresolved fluxes $\Theta_{0}^{m},\Theta_{k}^{m},Q_{k}^{M}$ at $t=t_{\ell}$ for the mean and variance using the LSTM model in \eqref{eq:lstm_homo} or \eqref{eq:lstm_inhomo}, {evaluated at the input parameter value $\theta$}, with the time-delay inputs from the previous $L$ time steps; 
\item Evaluate the perturbed mean states  $\hat{u}_{\delta,0}^M,\hat{u}_{\delta,k}^M$ at time $t_\ell$ using the mean models (the first two equations in \eqref{eq:model_ml_red3}). 
Subsequently, we attain the response mean statistics $\delta \hat{u}_0^M(t_\ell) := \hat{u}^M_{\delta,0}(t_\ell) - \hat{u}_{\textup{eq},0}$ and $\delta \hat{u}_k^M(t_\ell) := \hat{u}^M_{\delta,k}(t_\ell) - \hat{u}_{\textup{eq},k}$, where $\hat{u}_{\textup{eq},0},\hat{u}_{\textup{eq},k}$ are the reference equilibrium mean;
\item Update the effective damping and noise $D_{k}^{M},\Sigma_{k}^{M}$ at time $t_{\ell}$ using the decomposition in \eqref{eq:separation} of the statistical flux model $Q_{k}^{M}(t_{\ell})$; 
\item Update the stochastic coefficients $Z_{\delta,k}^{M,\left(i\right)}(t_{\ell})$ by solving the third equation in \eqref{eq:model_ml_red3} for each ensemble member with the effective damping and noise corrections.
\item Compute the empirical variances of the perturbed coefficients, $R_{\delta,k}^M(t_{\ell})=\frac{1}{M-1}\sum_{i=1}^{M}Z_{\delta,k}^{M,\left(i\right)}(t_{\ell})(Z_{\delta,k}^{M,\left(i\right)}(t_{\ell}))^*$. Subsequently, compute the response variance $\delta R_k^M(t_{\ell}):=R_{\delta,k}^M(t_{\ell}) - R_{\textup{eq},k}$, where $R_{\textup{eq},k}$  denotes the equilibrium variance of the unperturbed system;
\item Update $\ell = \ell +1$;
\end{itemize}
\EndWhile
\end{algorithmic}
\end{algorithm}

\section{Predicting leading-order statistics of the L-96 system\label{sec:Predicting-leading-order}}

In this section, we numerically validate the prediction skill of the proposed reduced-order statistical-stochastic models to recover the leading-order statistics on different statistical structures in the L-96 system. In particular, we consider two representative regimes, generating homogeneous and inhomogeneous statistics. The homogeneous regime provides a simpler test case for validating the proposed algorithm on {chaotic complex} systems with strong instability and non-Gaussian statistics. The inhomogeneous regime serves as a more challenging problem induced by nonlinear {spatio-temporal} interactions and non-zero cross-correlations. We organize the section as follows: First, we report the experiment configuration and the training data generation in Section~\ref{subsec:training_data}.  Then, we report the results for the homogeneous and inhomogeneous cases in Sections~\ref{subsec:homogeneous} and \ref{sec4.4}, respectively. 

\subsection{Model configuration and training dataset for the L-96 system}\label{subsec:training_data}

To generate the training data (or the label $y$ in \eqref{label}), we first integrate the L-96 system under  homogeneous reference forcing $F_{\mathrm{ref}}=8$ and
damping $d_{\mathrm{ref}}=1$. The equation is integrated using the 4th-order
Runge-Kutta scheme with a small time step $\mathrm{d}t=0.001$, and
the data is subsequently sampled at every 10 steps. Thus we have the data sampling
step $\Delta t=0.01$. The use of larger sampling step size, while introduce additional numerical discretization, is to reflect the practical situation when frequent measurements are not often available, especially if efficient numerical integration with larger time step is used. Subsequently, we compute the empirical mean and variance $\mathbf{\bar{u}}_{\textup{eq}}$ and $R_{\textup{eq},k}$, over these discrete time realizations. 

\begin{figure}
\begin{centering}
\includegraphics[scale=0.37]{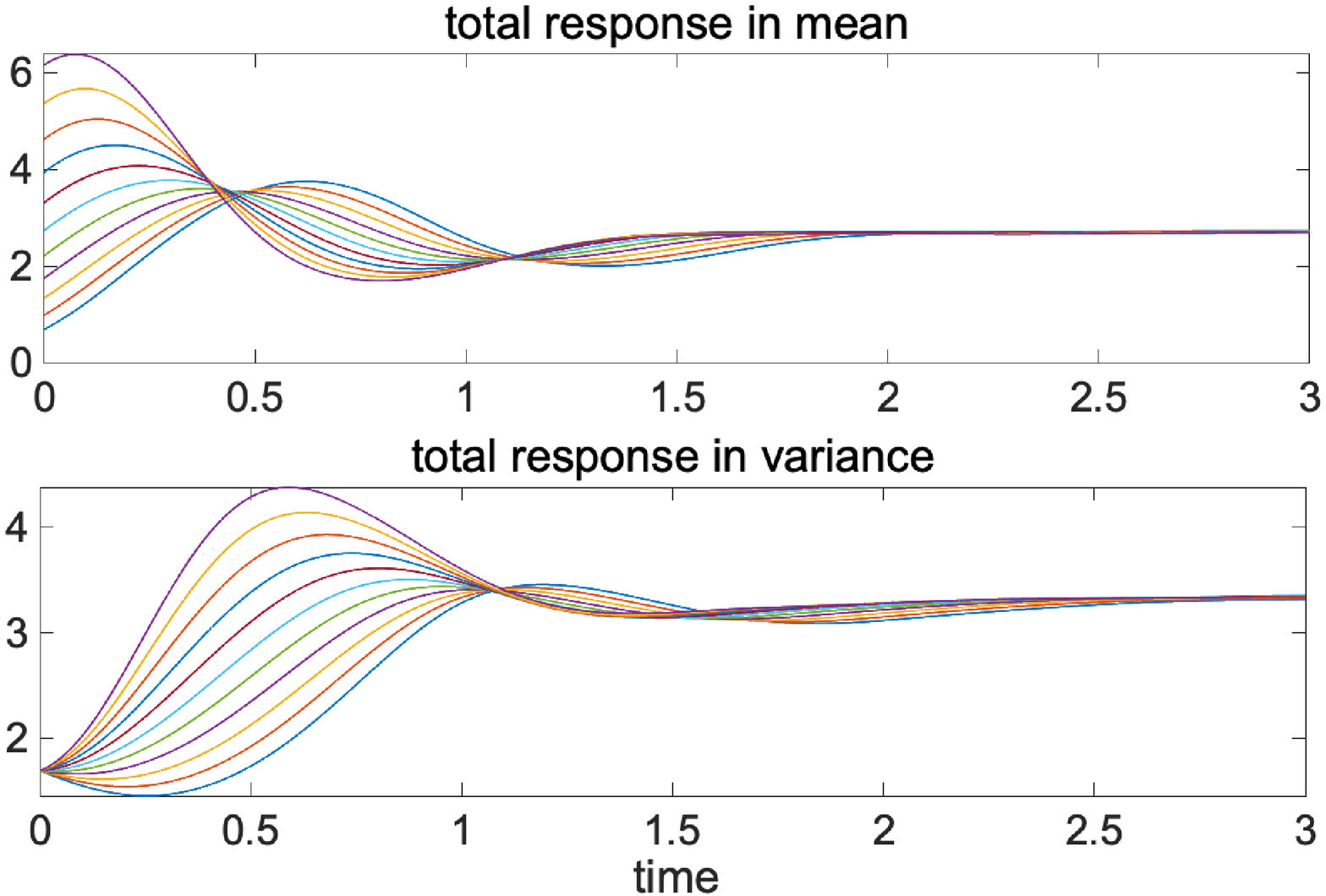}\includegraphics[scale=0.37]{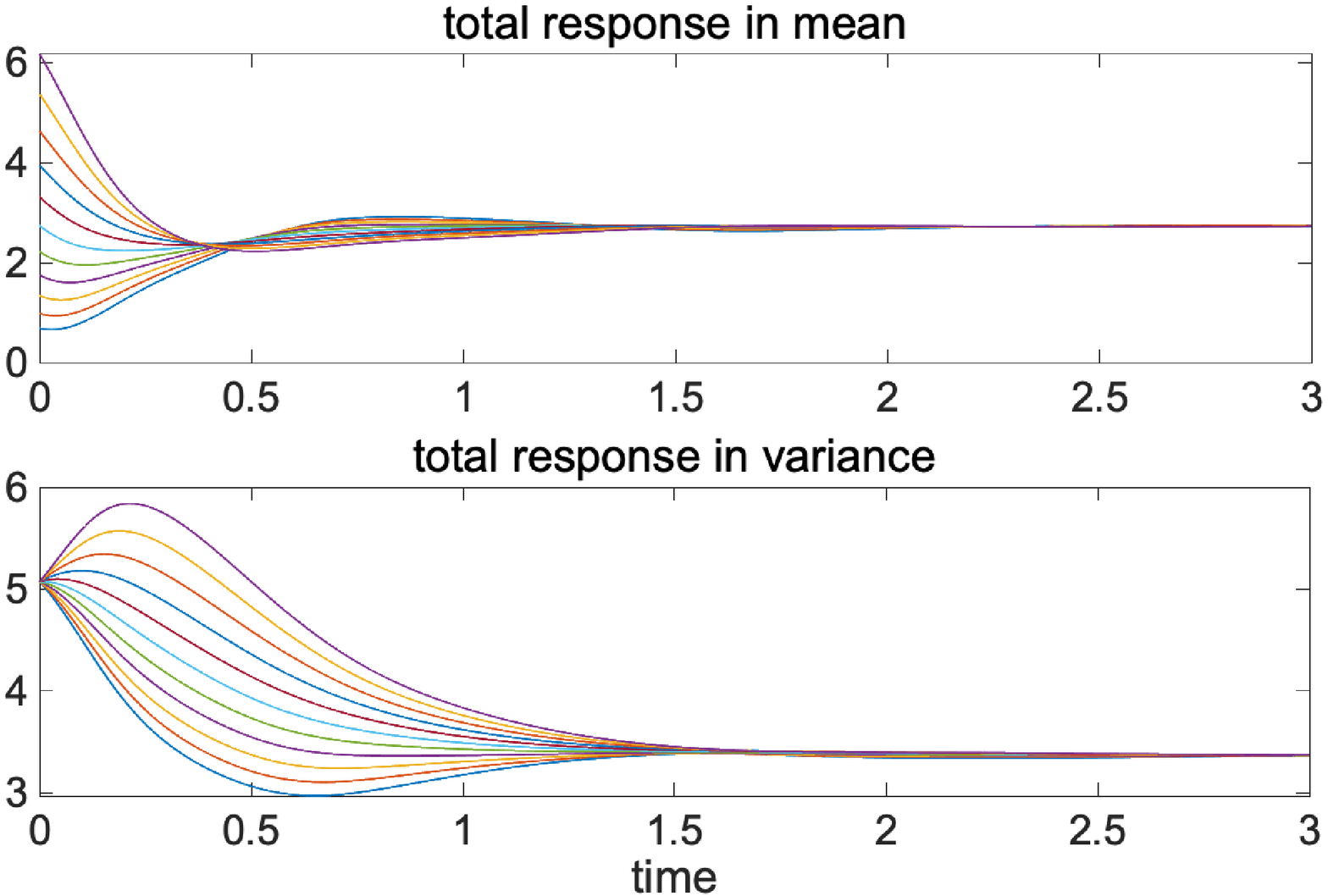}
\par\end{centering}
\caption{Statistical responses of the total energy in the mean and the total
variance of the L-96 system. Different lines represent the different
amplitudes of mean perturbation $\alpha\in\left[0.5,1.5\right]$.
Left: $\beta=0.5$; right: $\beta=1.5$.\label{fig:Responses}}
\end{figure}

To produce a unified training data set independent of the particular forcing and damping perturbations, we sample the transient state
statistics from only an initial perturbation of the ensemble samples
of the following form
\begin{equation}
\mathbf{u}^{\left(i\right)}:=\alpha\bar{\mathbf{u}}_{\mathrm{eq}}+\sqrt{\beta}\left(\mathbf{u}_{\mathrm{ref}}^{\left(i\right)}-\bar{\mathbf{u}}_{\mathrm{eq}}\right), \quad i=1,\ldots M = 500,\label{eq:samples}
\end{equation}
where $\{\mathbf{u}_{\mathrm{ref}}^{\left(i\right)}\}_{i=1,\ldots, M}$ denotes a set of {$M=500$ samples randomly drawn from a long discrete trajectory of solutions of the L-96 system corresponding to the reference damping and forcing $F_{\mathrm{ref}}, d_{\mathrm{ref}}$.}  New initial ensembles are generated by perturbing the mean through
the parameter $\alpha$ and the variance through the parameter $\beta$. Figure \ref{fig:Responses} plots several realizations of the statistical
responses for the mean and variance subject to different perturbation
parameters $\alpha,\beta$. The converging trajectories of the mean and variance also illustrate the decorrelation time that characterizes the mixing rate of the states. {The solutions will finally converge to the unperturbed equilibrium within the decorrelation time around $T=1.5$, which is a time scale that we expect the prediction skill to be accurate over the testing data.} For training, we will consider 6 different values for each $\alpha,\beta \in \left\{0.5, 0.7, 0.9. \ldots,1.5\right\}$, resulting to $6\times 6=36$ different initial perturbation cases.  

Based on these initial conditions, we have 36 trajectories of transient statistics for the reference systems (see some of these trajectories in Figure~\ref{fig:Responses}). To increase the number of training data in the homogeneous case, we consider 4 additional external constant forcings, $F_\delta = F_{\textup{ref}}+ \delta F$, where $\delta F \in \{-1, -0.5,0.5,1\}$ in addition to reference forcing with $\delta F = 0$. With these additional perturbations, we have $36\times 5=180$ trajectories of the response mean, $\delta \bar{\mathbf{u}}$, and variance statistics, $\{\delta R_k\}_{k\in\cI}$, at discrete time $t_\ell = \ell\Delta t\in [0, 2]$. For training data, we ignore the solutions beyond 2 time units since most of the statistical quantities at these times are constant. We partition the statistics on time interval $[0,2]$ into 10 overlapping sub-intervals, each of time length 1.1 units: $[0, 1.1], [0.1, 1.2], \ldots, [0.9,2]$. On each sub-interval, since the discrete time step is $\Delta t=0.01$, we have statistics at 111 data points. Following the notation in Pseudo-code~\ref{alg1}, we label these statistical timeseries as the quantities at $t_\ell = -L, \ldots, 0, 1, \ldots, 10$, with $L=100$. We will use the first $L+1 = 101$ data points of 
$\{\mathbf{u}^{(i)}\}_{i=1,\ldots,M}, \mathbf{\bar{u}}$ and $\{R_k\}_{k\in \cI}$  
to construct the input data $x$. Particularly, we take FFT on $\mathbf{\bar{u}}(t_\ell)\in\mathbb{R}^N$ to attain $\hat{u}_0(t_\ell)$ and $\{\hat{u}_k(t_\ell)\,:\, k\in\cI\}$ for $\ell = -100,\ldots, 0$. For the homogeneous case, it is clear that $\hat{u}_0 = \bar{u}$ and $\hat{u}_k = 0$ when $k\neq 0$. We also use the decomposition in \eqref{eq:mode_spec} on $\{\mathbf{u}^{(i)}(t_0)\}_{i=1,\ldots,M}$ to attain an ensemble perturbation $\{Z^{(i)}_k(t_0): k\in\cI, i=1,\ldots, M\}$. This completes the construction of an input $x$ for each partition. Finally, we take the last 10 data points of $\delta\mathbf{\bar{u}} (t_\ell)$ and $\{\delta R_k(t_\ell)\}_{k\in \cI}$ at $\ell=1,\ldots, 10$ as the label data $y$ on each partition.

\begin{figure}[ht]
\includegraphics[scale=0.37]{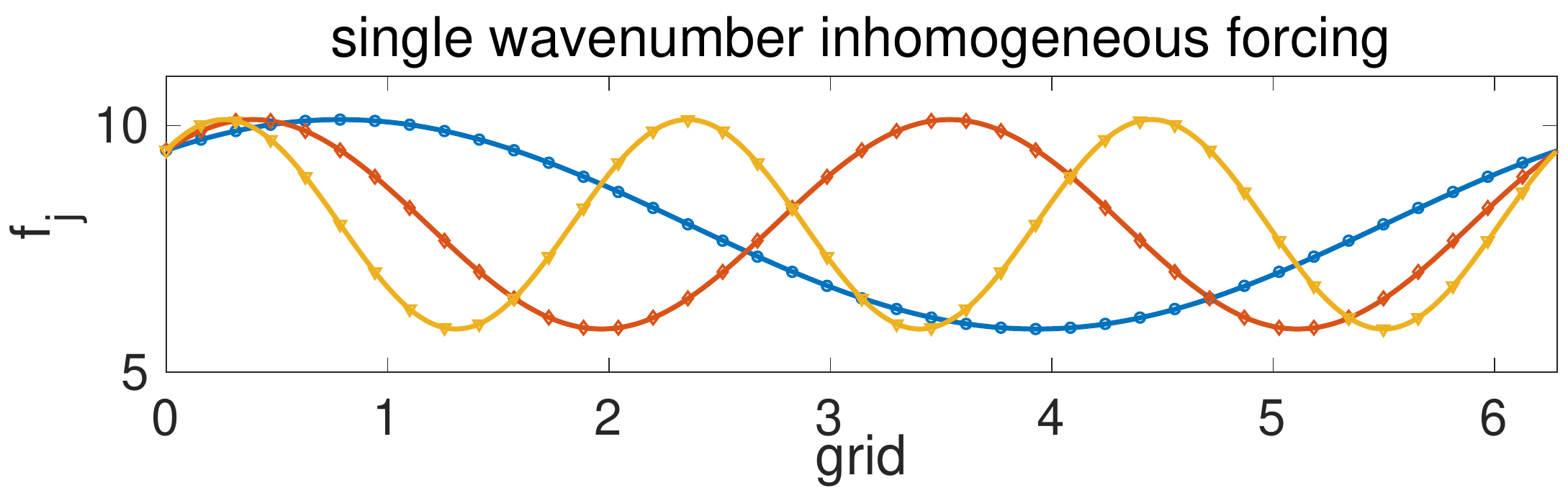}\includegraphics[scale=0.37]{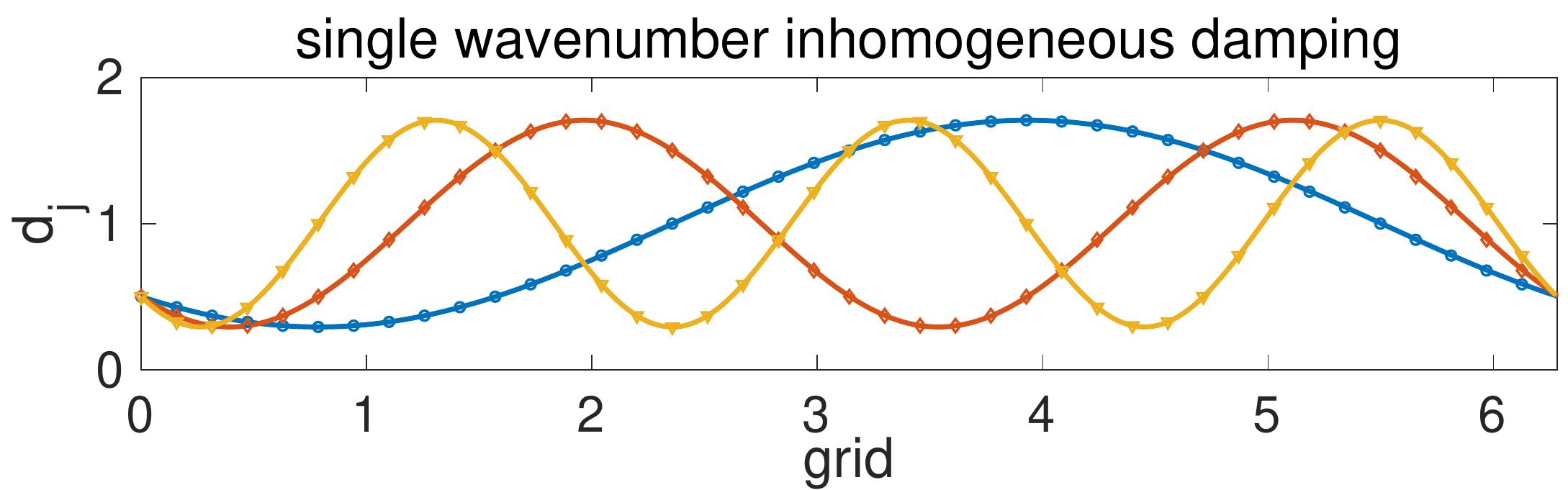}
\caption{Inhomogeneous forcing and damping perturbations on top of the equilibrium state $F_{\mathrm{eq}}=8$ and $d_{\mathrm{eq}}=1$. Perturbations are added to wavenumbers $k=1,2,3$.\label{fig:Different-types}}
\end{figure}

Accounting for the number of sub-interval from the partition, we have a total of $n=36\times 5\times 10 = 1800$ training data $(x_i,y_i)_{i=1,\ldots, n}$ for the homogeneous case. For the inhomogeneous case, we consider 16 different forcings and dampings, where in each case, the forcing is chosen to be one of the four cases: reference case $\delta f=0$ or {$\{\delta f_j = 1.5 \sin(\frac{2\pi kj}{N})\}_{k=1,2,3}$ and the damping is one of the four cases: reference damping $\delta d=0$ or $\{\delta d_j =0.5 \sin(\frac{2\pi kj}{N})\}_{k=1,2,3}$}. Effectively, these inhomogeneous forcings and dampings (see Figure~\ref{fig:Different-types}) exerted a single Fourier mode $k=1,2$ or $3$. Applying the same temporal partitioning as in the homogeneous case on each trajectory of response statistics, we have a total of $n=36\times 16\times 10=5760$ training data for the inhomogeneous case.

{
The training procedure is to solve the empirical risk minimization task of \eqref{empirical_risk}, where the average is over the training pairs $(x_i,y_i)_{i=1,\ldots, n}$ discussed above. In the empirical risk function in \eqref{empirical_risk}, the predicted label, $y_i^M =\mathcal{M}(x_i,\theta)$ is computed using Pseudocode~\ref{alg1}. Computationally, the optimization problem is to find the parameters $\theta$ in the LSTM models in \eqref{eq:lstm_homo} or \eqref{eq:lstm_inhomo} that is used in the first step of Pseudo-code~\ref{alg1}.} The hyper-parameters of the LSTM network are summarized in Table \ref{tab:Standard-model-hyperparameters}. We solve this optimization problem using the stochastic gradient descent (SGD) algorithm with batch size 1 to minimize the computational cost (as we do not find any advantage of using larger batch sizes). The learning rate is reduced by 50\% at iteration steps 25, 50, and 75. {Once the model is trained, denoting $\theta^*$ as the parameter obtained the from SGD algorithm, we use Pseudocode~\ref{alg1} to evaluate the response statistics, $\mathcal{M}(x_i^\mathrm{new},\theta^*)$, corresponding to new input $x_i^\mathrm{new}$ that is not in the training data set.}

\begin{table}[ht]
\centering
\begin{tabular}{cc}
\toprule 
total training epochs & 100\tabularnewline
\midrule 
ensemble size & 500\tabularnewline
\midrule 
SGD batch size & 1\tabularnewline
\midrule 
initial learning rate  & 0.001\tabularnewline
\midrule 
learning rate reduction at iteration step & 25, 50, 75\tabularnewline
\midrule
\midrule 
time step size between two measurements $\Delta t$ & 0.01\tabularnewline
\midrule 
LSTM sequence length $L$ & 100\tabularnewline
\midrule 
forward prediction steps in training $T$ & 10\tabularnewline
\midrule 
LSTM hidden state size $h$ & 100\tabularnewline
\bottomrule
\end{tabular}

\caption{Hyper-parameters for training the standard {Long-Short-Term-Memory (LSTM) neural network model using stochastic gradient descent (SGD)}.\label{tab:Standard-model-hyperparameters}}
\end{table}

\subsection{Training and prediction of the homogeneous statistical regime}\label{subsec:homogeneous}

First, we consider the homogeneous perturbation case using uniform perturbations in the forcing $\mathbf{F}=F_{\mathrm{eq}}+\delta f\mathbf{e}_{0}$. Only the most energetic leading modes, $\cI = \{k\,:\,6\leq\left|k\right|\leq12\}$, are resolved in the fluctuation equation for $Z_{k}^M$ (compared to total 40 modes). Here, the target is to predict the homogeneous mean state $\bar{\mathbf{u}}=\bar{u}\mathbf{e}_{0}$
and the diagonal variance in resolved mode $R_{k}=\frac{1}{M-1}\sum_i Z_k^{(i)}(Z_k^{(i)})^*$ based on the ensemble solutions.

\begin{figure}
\centering
\subfloat[Errors in mean and total variance during training iterations]{\includegraphics[scale=0.5]{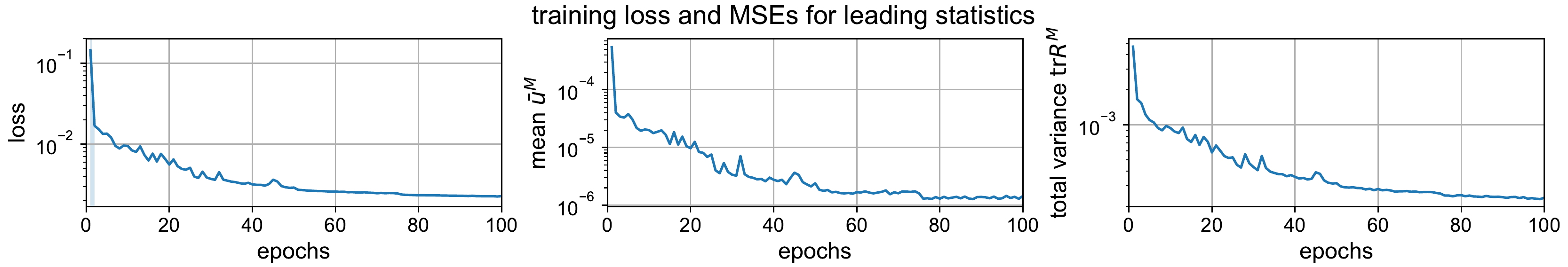}
}

\subfloat[Errors in the unresolved fluxes during training iterations]{\includegraphics[scale=0.5]{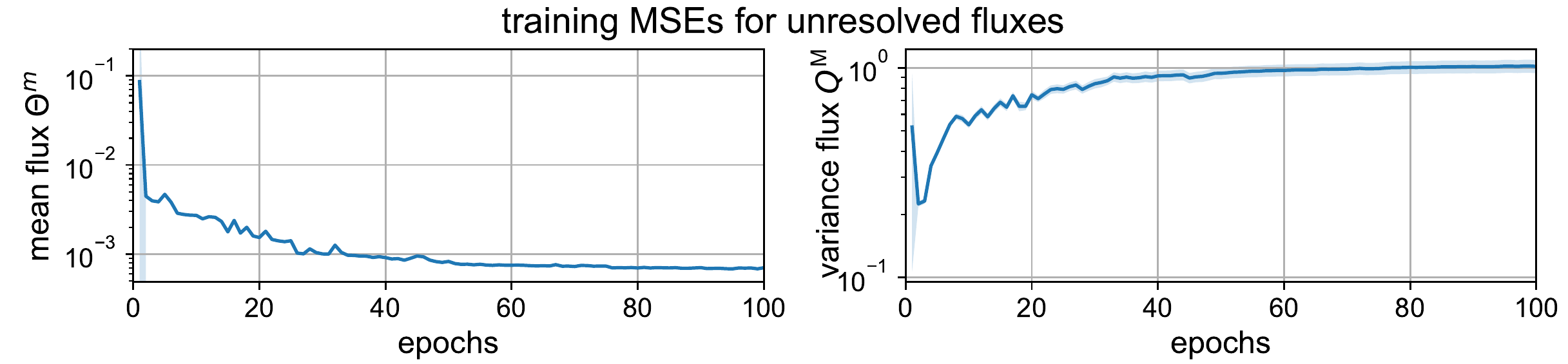}
}

\caption{Training errors using the reduced-order model (\ref{eq:model_ml_red2})
for the homogeneous statistical regime. The first row shows the evolution of the loss function and MSEs in the predicted mean and variance. The second row compares the difference between the true flux and the neural network outputs in mean $\Theta^m$ and $Q^M_k$.\label{fig:Training-homo}}
\end{figure}

We first show the evolution of the errors during the training iterations in Figure \ref{fig:Training-homo}(a). The first row plots the values of the empirical risk function \eqref{empirical_risk}, where the training errors in the predicted mean, $\bar{u}^M$, and variance, $R_k^M$, are computed based on the empirical average of the ensemble of solutions with training inputs, $\{x_i\}_{i=1}^n$. The neural network model is trained with 100 repeating epochs and a small number of forwarding steps $T=10$. The result shows that the loss function can be minimized to small values after a much smaller number of iterations (around 40 epochs). Correspondingly, the {mean square errors ($\mathrm{MSE}(f,f^M)=\frac{1}{n}\sum_{i=1}^n|f_i-f^M_i|^2$) of the homogeneous mean $\bar{u}$} and the total variance of resolved modes $\mathrm{tr}R^{M}=\sum_{k\in\cI}R_{k}^{M}$ can be both effectively minimized to very small values. {This indicates the accurate fitting of both the mean and statistics in the reduced order model.} The variance of the sample errors is also plotted by the shaded area around the lines, which is reduced to negligibly small values. The decay of training error demonstrates the effectiveness of the training process in reducing the model errors uniformly among all the training samples. For more detailed comparisons of the training performance, we also plot the training errors in the neural network outputs of the unresolved flux terms $\Theta^m$ and $Q^M_k$ in \eqref{eq:lstm_homo} that are not directly compared in the loss function. In this case, we found that the error in $\Theta^m$, which is not measured directly in the loss function, decays. On the other hand, the discrepancy between the model constructed flux, $Q^M_k$, and the statistical flux, $Q_{F,k}$, actually increases. This is not a surprise since we used the decomposition in \eqref{damping_noise_decomposition} to determine $D_k^M$ and $\Sigma_k^M$.  Recall that this parameterization uses the equilibrium variance $R_{\mathrm{eq},k}$ to avoid the elaborate computational cost induced by fitting to the more ideal time-dependent variance, $R_{k}$, as suggested in \eqref{damping_noise_decomposition}.

\begin{figure}
\centering
\includegraphics[scale=0.6]{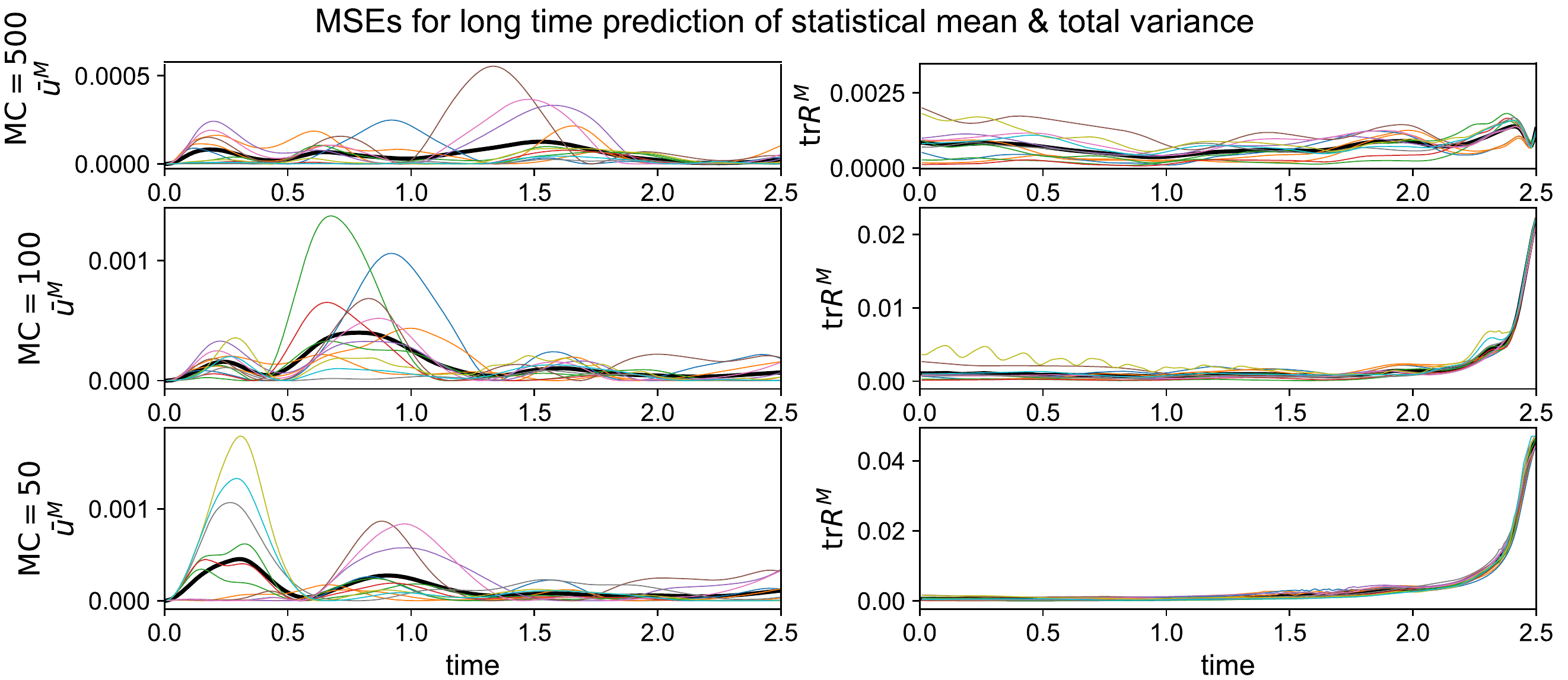}

\caption{Prediction errors using the trained reduced-order model (\ref{eq:model_ml_red2})
for the homogeneous statistical regime. MSEs of initial conditions that are different than the training input data (in thin colored lines) in the mean state and ensemble variance are compared together with the overall average of all the test cases (in thick black line). Errors using different ensemble sizes $\mathrm{M}=500,100,50$ to
recover the statistics are also compared.\label{fig:Prediction-err-homo}}
\end{figure}

In the forecast stage, the trained model is applied to predict
the key statistics under perturbations of initial conditions that are different from the training input data. We should point out that the model output in the forecast stage at time $t_i$ is the result of iterating the model $i$ steps, thus the model errors are accumulated through these iterations, and attaining accurate prediction becomes challenging for the unstable modes with a positive growth rate as illustrated in Figure \ref{fig:instability}.  The prediction errors in the mean state and variance from the empirical ensemble average are plotted in the first row of Figure \ref{fig:Prediction-err-homo}. The errors in the test solutions with different perturbations of initial conditions are compared during the time evolution up to a long time $T=2.5$ with $250$ iterations ({beyond the decorrelation time $T_\mathrm{decorr}\approx1.5$ of the state}). The result suggests that the trained model produces accurate statistical predictions under various tested perturbations of initial conditions.  Particularly, the errors in both mean and variance in resolved fluctuation modes remain small during the prediction time interval shown. This confirms the {stable model dynamics} using the effective damping and forcing introduced in this reduced model. 

\begin{figure}
\centering
\includegraphics[scale=0.55]{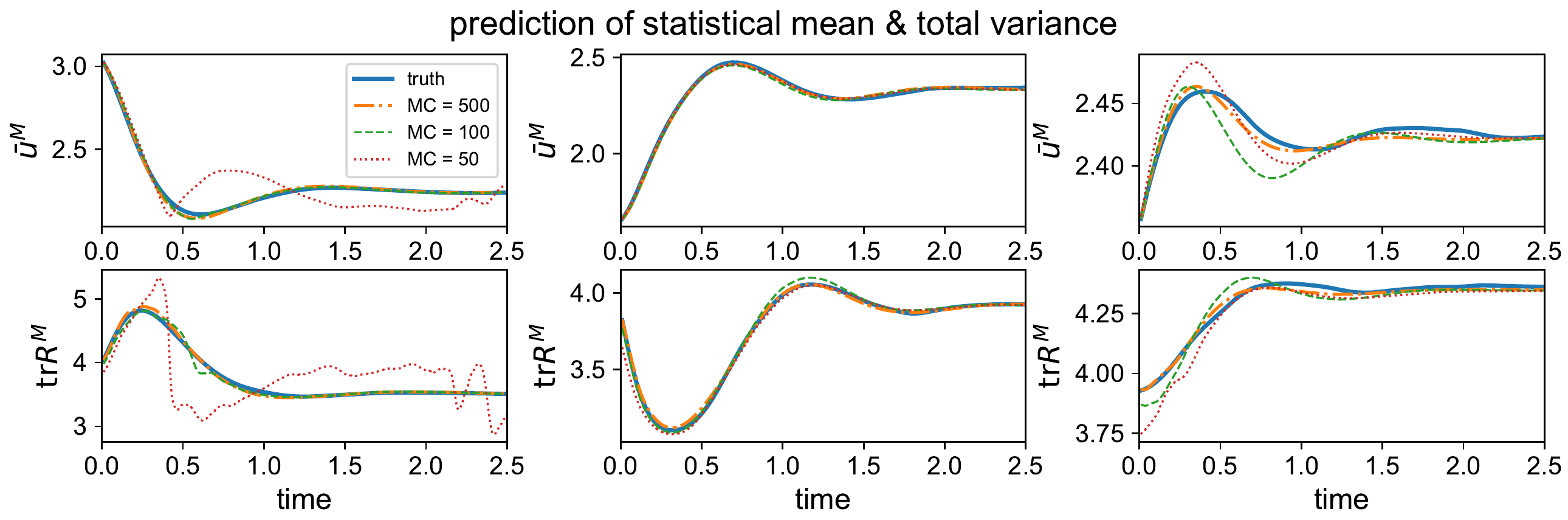}

\caption{Prediction of the statistical mean and variance using the trained
reduced-order statistical-stochastic model for homogeneous statistics. The predictions of the mean and total variance with different ensemble sizes $M=500,100,50$ are compared. The first, second and third columns show predictions starting from an initial condition that does not belong to the training input data under three forcing perturbations $\delta F = -1, 0, 1 $ that were used to generate the training data set, respectively. \label{fig:Prediction-homo}}
\end{figure}

We also report the prediction skill for the smaller number of samples in the second and third row of Figure \ref{fig:Prediction-err-homo}. Specifically, the prediction errors using smaller ensemble sizes $M=100, 50$ are compared using the same model that is trained with an ensemble of size $M=500$. Compared with the larger ensemble case $M=500$ in the first row, the errors begin to grow as the ensemble size decreases. This is expected since the estimated statistics via ensemble average become less accurate. However, we can still see that the prediction is accurate in most of the test cases. A more detailed comparison between the statistical predictions of the mean, the trace of the variance, and the variance of each resolved mode under three forcing perturbations are shown in Figure \ref{fig:Prediction-homo}. Consistent with the prediction errors in Figure \ref{fig:Prediction-err-homo}, the evolution of the mean and variances starting from three {pairs} of initial conditions and forcings is captured accurately when $M=500$. 
When the ensemble size is reduced to $M=100$, there is a slight increase in errors, however, the overall qualitative transient behavior in each resolved mode is still accurately predicted. When the ensemble size is reduced to $M=50$, the performance accuracy varies wildly. Under large forcing perturbations $\delta F = \pm 1$ (see the first and third columns), the prediction accuracy significantly deteriorates. Under the reference perturbation with $\delta F =0$ in the second column, the statistics are accurately predicted. 

\subsection{Training and prediction of the inhomogeneous statistical regime}\label{sec4.4}

Finally, we consider the model prediction skill in the challenging
case induced by inhomogeneous statistics. As we have discussed in Section~\ref{subsec:training_data}, we train the model using data generated by applying inhomogeneous damping and forcing on the first three leading modes, which corresponds to spatially periodic forcing and damping corresponding to these wave numbers as shown in Figure~\ref{fig:Different-types}.

\begin{figure}
\centering
\subfloat{\includegraphics[scale=0.45]{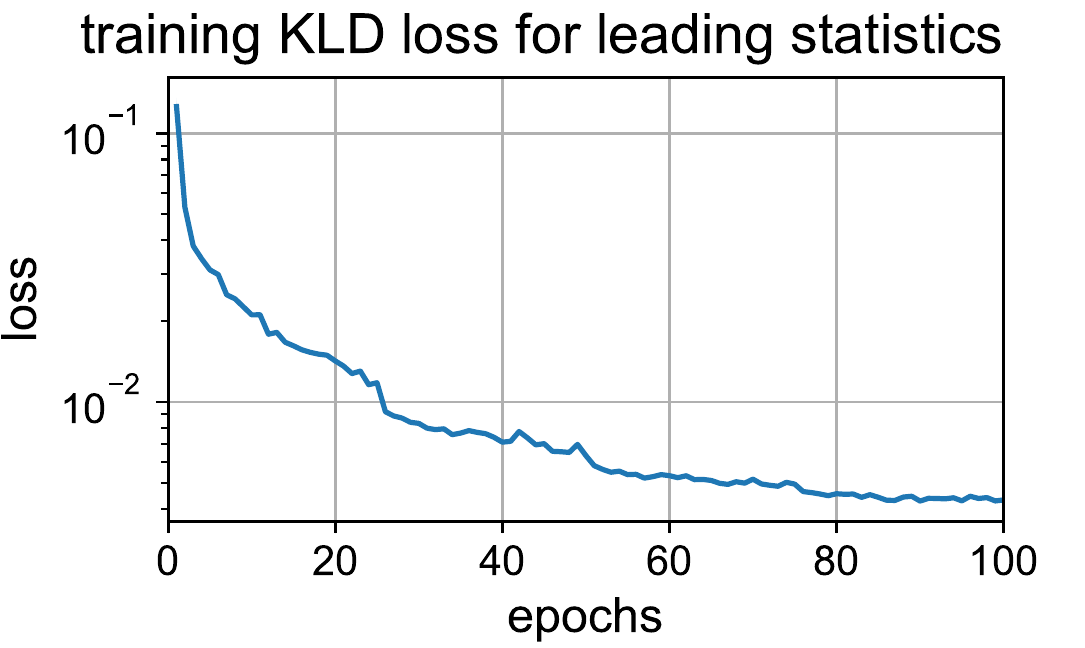}\includegraphics[scale=0.45]{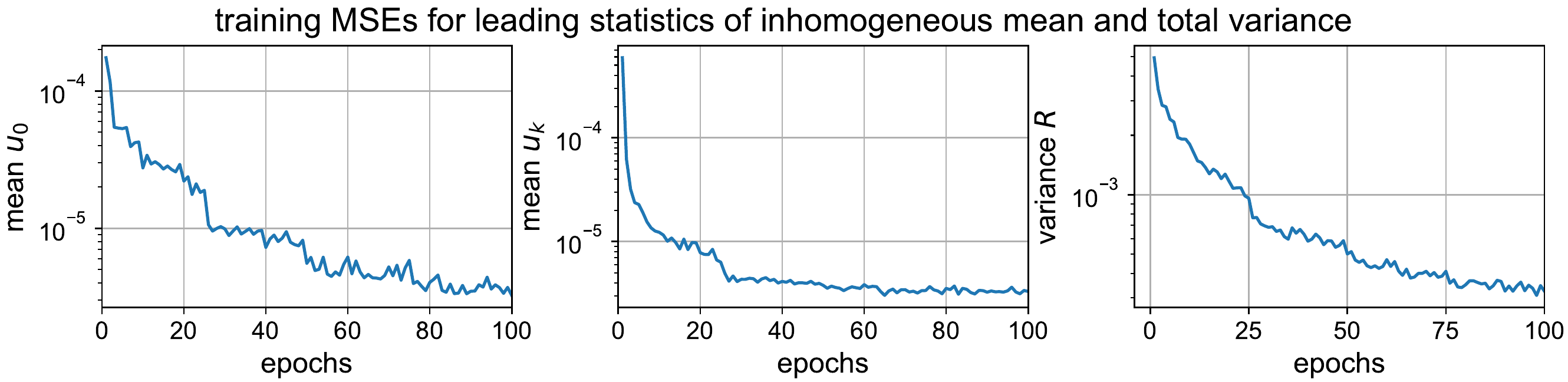}}

\subfloat{\includegraphics[scale=0.45]{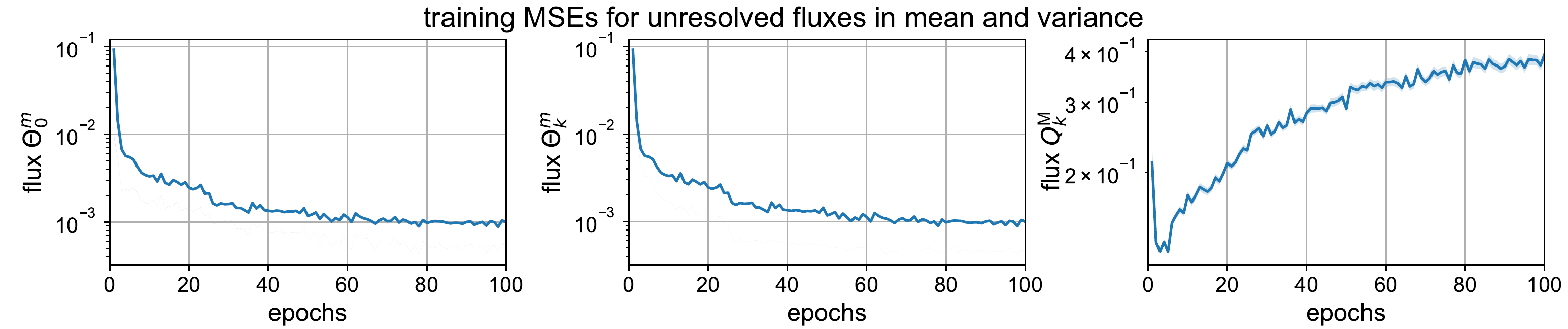}}

\caption{Training relative entropy loss and MSEs during training iterations of the reduced stochastic model.\label{fig:MSEs-during-training} }
\end{figure}

As in the previous section, we first display the training results using the reduced-order model (\ref{eq:model_ml_red3}) to learn and recover the inhomogeneous statistics of the perturbed L-96 system. In this inhomogeneous case, the resolved mean modes include the inhomogeneously forced and damped wavenumbers $k=1,2,3$, and the resolved fluctuation modes include still the most energetic ones $6\leq\left|k\right|\leq12$. The evolution of the errors during training iterations is displayed in Figure \ref{fig:MSEs-during-training}. 
Similar to the homogeneous case, the errors can be effectively minimized within the 100 training epochs in both the homogeneous mean $\hat{u}_0 = \bar{u}$ and variance $R_{k}$ as well as all the inhomogeneous mean Fourier coefficients $\hat{u}_{k}$. Again, it is useful to notice that the error in the variance feedback $Q_{k}^{M}$ actually increases during the training process. As in the homogeneous case, this discrepancy is due to the use of coefficients $D_k^M$ and $\Sigma_k^M$ attained by equilibrium fitting in \eqref{eq:separation} as a way to realize the decomposition in \eqref{damping_noise_decomposition}.

\begin{figure}[ht]
\subfloat[prediction MSEs in different test cases]{\includegraphics[scale=0.6]{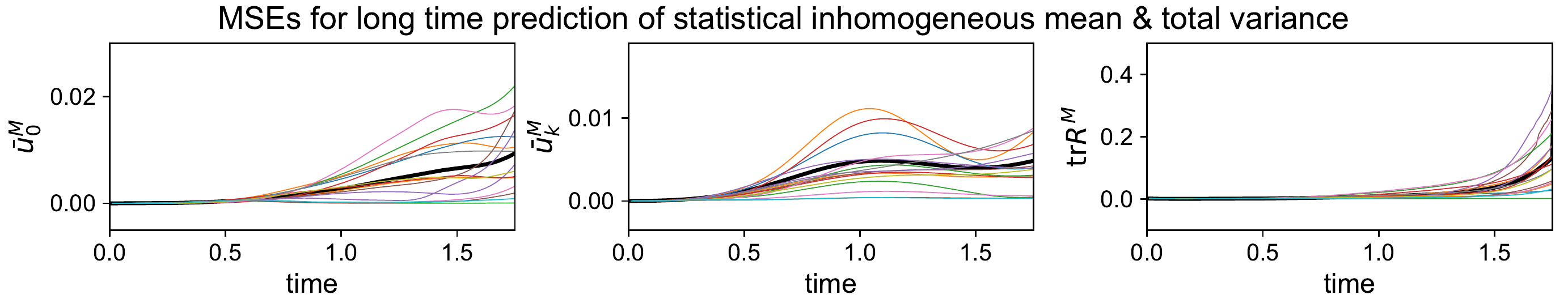}}

\subfloat[prediction of the inhomogeneous statistics in 3 initial perturbation cases (in 3 columns)]{\includegraphics[scale=0.6]{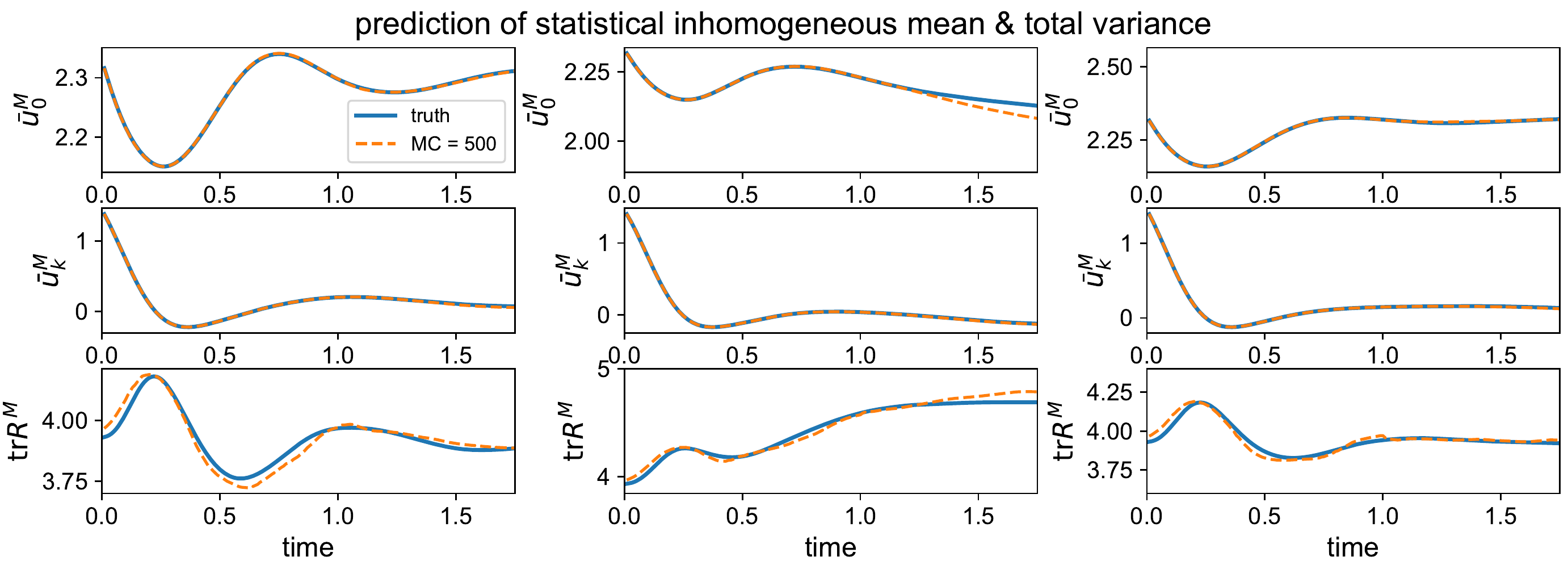}}

\caption{Upper panel: Prediction errors in the homogeneous mean $\bar{u}_0^M$, resolved inhomogeneous mean $\bar{u}_k^M=\sum_{k\in\mathcal{I}}\hat{u}_k^M$ and total variance $\mathrm{tr}R^M=\sum_{k\in\mathcal{I}}r_k^M$ with inhomogeneous
forcing and damping. Errors with different initial perturbations are plotted in thin colored lines and the total averaged error is plotted in thick black
lines. Lower panel: Prediction of the homogeneous and resolved inhomogeneous mean, and total resolved variance for the same initial state perturbation that is not in training data set.  Each column corresponds to a specific choice of damping and forcing that is used to generate the training data set. The true solution is in solid blue line and the model prediction is in dashed orange line.\label{fig:Prediction-errors-samps} }
\end{figure}

\begin{figure}
\centering
\subfloat[inhomogeneous mean in each resolved mode ]{\includegraphics[scale=0.5]{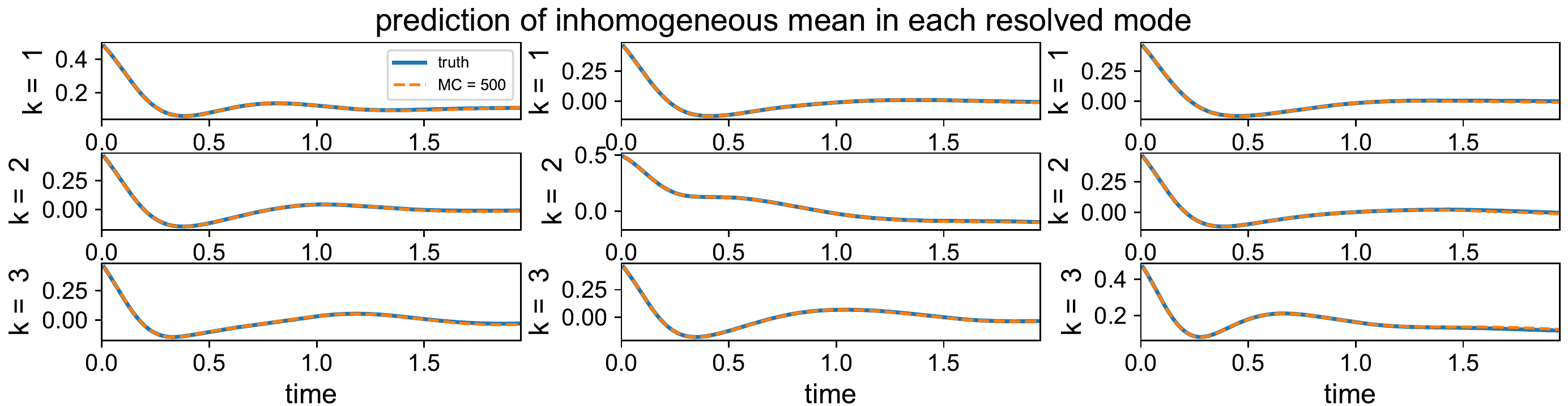}}

\subfloat[variance in each resolved mode ]{\includegraphics[scale=0.52]{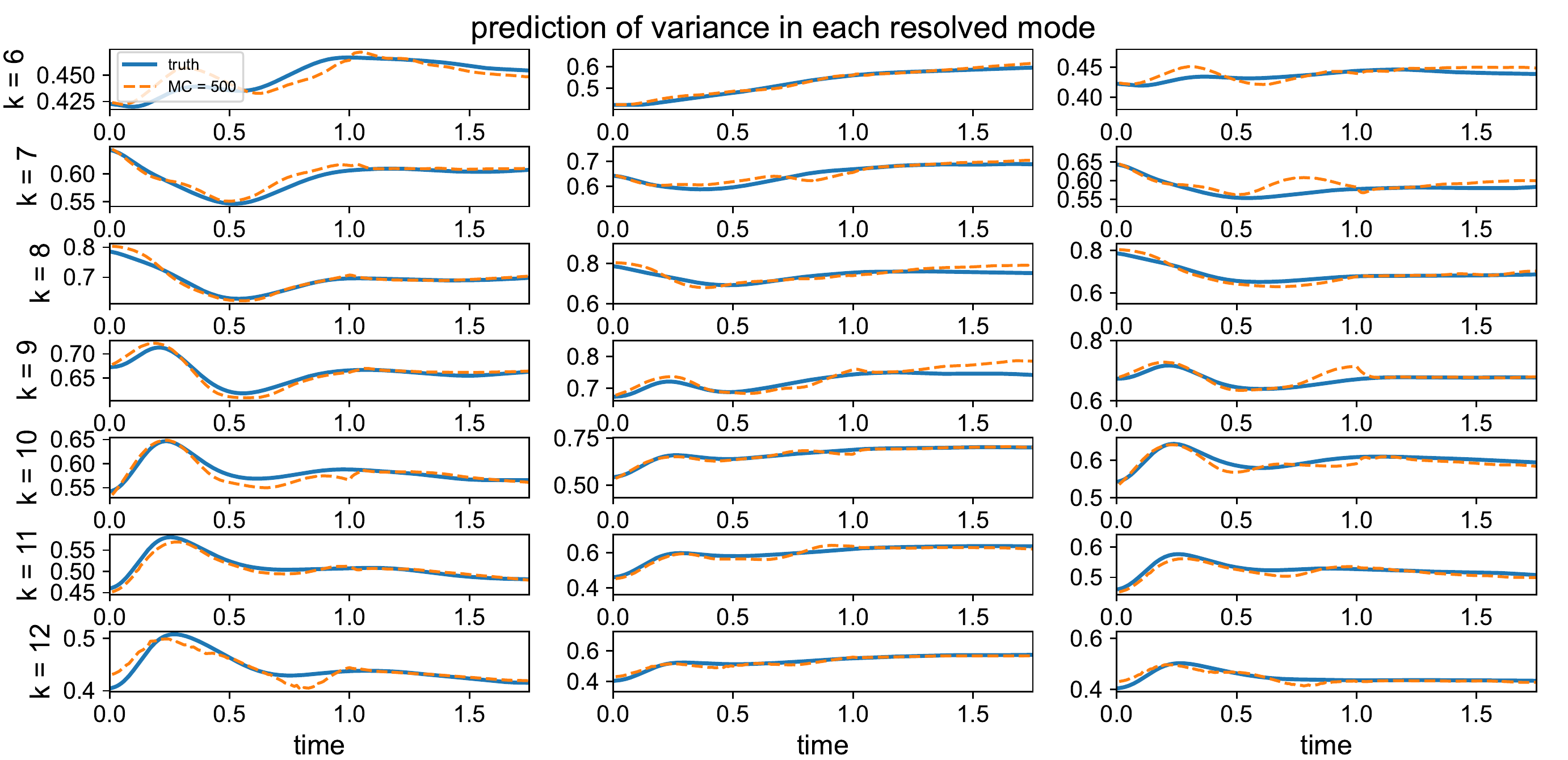}}

\caption{Detailed prediction results of the statistical mean state in the first three inhomogeneous modes, and the predicted variance in each resolved mode. The three columns represents 3 typical test cases with different perturbations.\label{fig:Detailed-prediction}}
\end{figure}

Next, we check the long-term prediction of the trained model for capturing the inhomogeneous mean and variance, starting from new initial conditions that are different from the training input data. As in the homogeneous case, instead of iterating the model in a small number of steps (10 forward steps) in the training stage, the prediction stage iterates the optimized model in 175 steps to achieve prediction up to $1.75$ model unit time. 
The first row of Figure \ref{fig:Prediction-errors-samps} shows the prediction MSEs in the homogeneous and inhomogeneous mean state and resolved variance under three inhomogeneous forcing and damping cases on wavenumbers $k=1,2,3$ as in Figure~\ref{fig:Different-types}, for new perturbations on the initial mean and sample variance as in \eqref{eq:samples}.
This is to check the model responses in leading order statistics subject to the inhomogeneous statistical structures induced by the forcing and damping. It shows that the trained reduced-order model produces accurate prediction skill on both the homogeneous and inhomogeneous components among all the different test cases.
The next three rows of Figure \ref{fig:Prediction-errors-samps} display the detailed comparison of the true and predicted statistics corresponding to the same initial condition for three different damping and forcing perturbations (that are imposed to obtain the MSE in the first row). The predicted inhomogeneous mean state in the first three modes and responses in leading variance mode are also compared in Figure \ref{fig:Detailed-prediction}
for the three different perturbation cases.

\begin{figure}
\centering
\subfloat{\includegraphics[scale=0.42]{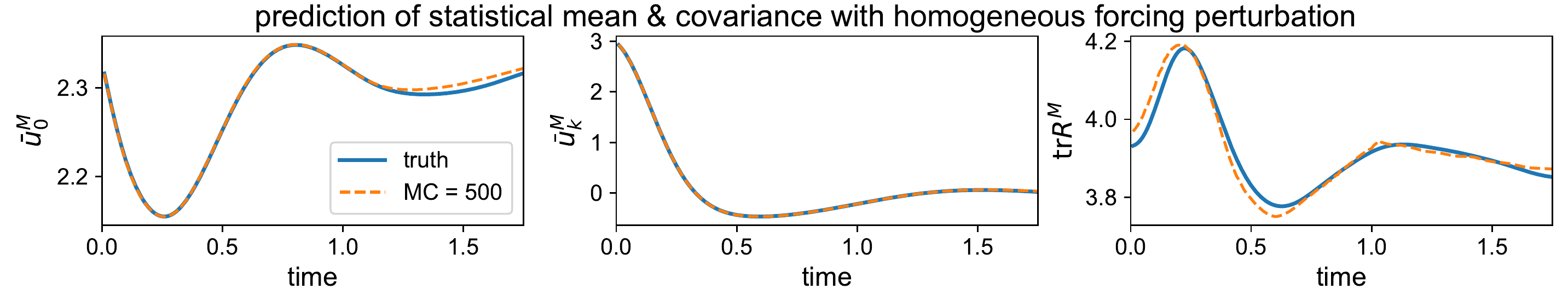}}

\subfloat{\includegraphics[scale=0.42]{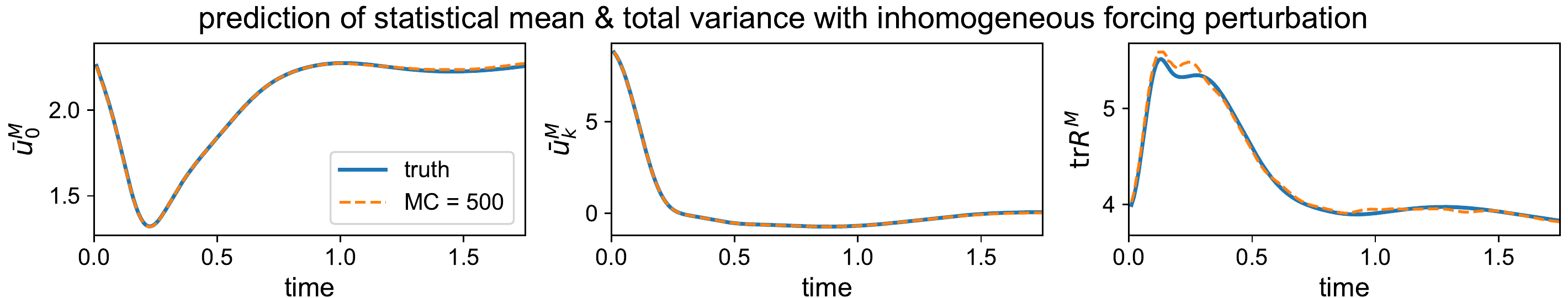}}

\caption{Prediction of the mean and total resolved variance with external forcing perturbation on the mean $\delta f = 0.5+1.5\sin(x)$ (upper) and on the first 3 inhomogeneous modes $\delta f= f_1+f_2+f_3$ (lower) with inhomogeneous statistics. The true solution is in solid blue line and the model prediction is in dashed orange line.\label{fig:Prediction-forcing}}
\end{figure}

While all previous numerical experiments were focused to capture statistical responses under new initial state perturbations, we further test the model prediction for mean and variance responses to different external forcing perturbations. The additional forcing perturbations are exerted on either the homogeneous mean state {$\delta f = 0.5+1.5\sin(\frac{2\pi j}{N})$} or the inhomogeneous leading mean modes $\delta f = f_1+f_2+f_3$, with {$f_k = \sin(\frac{2\pi kj}{N})+\cos(\frac{2\pi kj}{N})$}, where these additional constant mean forces are not in the training data set. Figure \ref{fig:Prediction-forcing} shows that the closure model predicts the statistical responses accurately under these different forcing perturbations. 

We should point out that for inhomogeneous cases, the inhomogeneous mean and cross-covariances play an important role. Thus the 
the dynamics of covariance in \eqref{eq:cov_dyn} is fully coupled with the inhomogeneous mean modes. This fact makes accurate dynamical modeling of even just the variance components nontrivial since one has to account for the interaction with all other modes. {In addition, the inherent instability in the dynamical model will amplify the unavoidable small errors in the neural network output at each time iteration step and accumulate them in time.} Despite these challenging issues, we found that the proposed reduced-order model can accurately predict the response mean and variance statistics under different initial and forcing perturbations for a  {long prediction time beyond the decorrelation time of the states} before the error starts
to accumulate in time. 
 
\section{Summary}\label{sec:Summarizing-discussions}
In this paper, we proposed a generic statistical-stochastic closure modeling framework for effective ensemble prediction of leading order statistics in complex systems containing strong instability and {interactions among different spatio-temporal scales}. The mean dynamics are modeled with a set of statistical equations that represents the homogeneous and inhomogeneous components subject to external perturbations. The fluctuation dynamics, which characterize the uncertainty among the multiscale modes, are modeled with a stochastic formulation. The mean equations are coupled with the covariance matrix that is empirically estimated using the ensemble prediction of the fluctuation terms. On the other hand, the stability of the fluctuation dynamics depends on the mean state. Such a formulation guarantees the positive definiteness of the covariance matrix. A reduced-order closure strategy is formulated to resolve the most energetic mean and variance modes for efficient computation with an ensemble simulation. Subsequently, machine learning tools are adopted to identify non-Markovian models for the nonlinear feedback from unresolved processes and imperfect model errors. 

{To combat instability and allow for a scalable training procedure, we considered a closure model with an effective damping and noise parameterization of the form \eqref{eq:flux_decomp}. Here, the damping coefficients and noise amplitudes are identified with a non-Markovian model that is designed to be consistent with the unperturbed equilibrium covariance statistics in the long term. Further diagonal approximation is employed in our numerical examples where the covariance statistics are diagonally dominant. Finally, efficient training of the neural network model under limited training dataset is achieved by measuring only the statistical output using an information metric-based loss function, fitting to the available training response mean and variance statistics. With such a training procedure, we effectively avoid the usually exhausting process of fitting of a large number of high-dimensional stochastic trajectories that often leads to overfitting as illustrated in \ref{appen:Predicting-leading-order-statist}. 

We numerically found that the proposed approach is effective in identifying stable dynamics with accurate statistical prediction. The dimension of the reduced-order model is of order $K \left(1+M\right)$, where the resolved state dimension $K\ll N$ is much smaller than the dimension of the full state, $N$. Thus, a smaller ensemble size $M$ is needed to sample the lower dimensional resolved subspace. The skill of the reduced-order statistical-stochastic model is tested on several statistical regimes of the L-96 system, including the homogeneous and inhomogeneous statistics produced by exerting different types of forcing and damping perturbations. In the simpler homogeneous case, the trained model showed uniformly high skill in dealing with different perturbations. In this case, the reduced-order model only resolves the first seven most energetic complex-valued leading Fourier modes. In the inhomogeneous case, the situation becomes more challenging. Inhomogeneous perturbations exert the mean energy of some Fourier modes with low variance energy and the covariance becomes non-diagonal although it is still diagonally dominant. Including three additional Fourier modes corresponding to the largest mean energy in the reduced-order model, we achieve accurate predictions of the leading mean and variance states on various nontrivial inhomogeneous statistical regimes and obtain a model that remains stable for at least up to the decorrelation time of the states. However, we should also point out that there is no guarantee that the present model does not blow up if we keep iterating for longer times since the closure parameterization ignores the non-diagonal component of $Q_F$ which may suppress conditional linear instability. 

While the numerical results are encouraging, constructing an improved parameterization that accounts for non-diagonal components of the unresolved feedback $Q_F$ that can guarantee stable dynamics for a long time under various perturbations induced by the learning and the temporal discretization errors remains an open issue. While finding a stable parameterization for $Q_F$ is a general problem, we also believe that this issue is related to an open problem in linear response theory for chaotic dynamical systems. The study of the validity of linear response in \cite{baladi2014linear} is crucially important to predicting the response at the steady state of the perturbed dynamical system. Particularly, the main goal of such a study is to check whether the perturbed system has an invariant measure that is smooth (under appropriate topology) as a function of the parameters that reflect the perturbations. If such a condition is invalid, then small perturbations induced by the learning error of the approximate closure model will generate a drastic change in the dynamical behavior, and thus, prohibit us to emulate accurate long-time statistics.
}

{
\section*{Data availability}
The codes are written in Python and are available on GitHub (https://github.com/qidigit/non-Markovian-closure-LSTM).}

\section*{Acknowledgments}
The research of D.Q. was partially supported by the start-up funds and the PCCRC Seed Funding provided by Purdue University. The research of J.H. was partially supported by the NSF grants DMS-1854299, DMS-2207328, DMS-2229435, and the ONR grant N00014-22-1-2193.

\appendix
\renewcommand\theequation{A\arabic{equation}} 
\setcounter{equation}{0}
\renewcommand\thefigure{A\arabic{figure}}     
\setcounter{figure}{0}
\section{Trajectory training and prediction using the direct stochastic model\label{appen:Predicting-leading-order-statist}}

In this Appendix, we demonstrate the limitation of the standard learning procedure with a loss function that measures the discrepancies between stochastic trajectories. Since the reduced-order model couples a statistical quantity that depends on empirical variance of a stochastic fluctuation, fitting trajectories of an entire statistical-stochastic system (such as \eqref{eq:model_ml_red1}) can be numerically demanding, especially when $M$ is large.  In the following, we will conduct an experiment fitting only the stochastic component of \eqref{eq:model_ml_red1} by assuming that the time series of the underlying mean $\bar{u}$ is always available for us, and thus, ignoring the error induced by finite ensemble size $M$ in the mean dynamics. While this scenario is not useful for real-time prediction, we will demonstrate that the standard machine learning procedure may not produce an effective learning even in such a simple case when the proposed damping and forcing parameterization in \eqref{damping_noise_decomposition} is not used.

Specifically, we parameterize $\Theta_k^v$ in \eqref{eq:model_ml_red1} by minimizing an empirical risk defined with the following loss function,
\begin{equation}
\textup{L}(\theta,\mathbf{Z}^\cI,\mathbf{Z^M}): = \sum_{k\in\cI} | Z_k - Z_k^M |^2, \label{newloss}
 \end{equation}
 where, 
 \begin{eqnarray}\label{partialsystem}
 \begin{aligned}
 \frac{dZ_{k}^{M}}{dt} & = -\left(\gamma+\gamma_{k}\bar{u}\right)Z_{k}^{M}+\Theta_{k}^{v}, \\
 \Theta_{k}^{v} (t_{i+1}) &=   \Theta_{k}^{v}(t_i) +  \mathrm{LSTM}^{m}\left(\bar{u}\left(t_{i-L:i}\right),\left\{ R_{k}\left(t_{i-L:i}\right)\right\} ,\Theta_k^{v}\left(t_{i-L:i}\right);\theta\right).
 \end{aligned}
 \end{eqnarray}
 In this numerical experiment, the input data is
 \[
 x = \left(Z^M_k(t_i), \bar{u}(t_{i-L:i}),R_{k}(t_{i-L:i}),\Theta_k^{v}(t_{i-L:i})\right) \in \mathbb{R}^{(2K + 1)L + K}.
 \]
 and output is $y = \mathbf{Z} \in \mathbb{R}^{K}$. In this homogeneous statistics case, we set $\cI=\{k:6\leq |k|\leq 12\}$ as in Section~\ref{subsec:homogeneous}, which results in $K=|\cI|= 14$. Setting $L=100$ as in Table~\ref{tab:Standard-model-hyperparameters}, this learning problem is to find a $(2K + 1)L + K = 2914$ dimensional map, which is quite high-dimensional. We fit this into the fluctuation coefficients $\{Z_k:k\in\cI\}$ corresponding to the statistical training data for homogeneous case discussed in Section~\ref{subsec:training_data}. Since we are fitting each realization of the fluctuation, the size of training data set is $n=1800\times 500$, accounting $M=500$ ensemble members.

\begin{figure}[ht]
\centering
\subfloat[loss and MSEs during training iterations]{\includegraphics[scale=0.5]{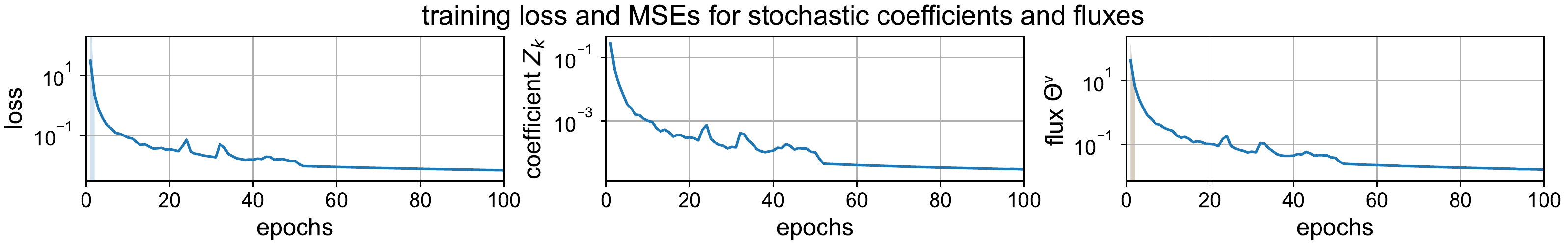}}

\subfloat[trajectory prediction using training data]{\includegraphics[scale=0.55]{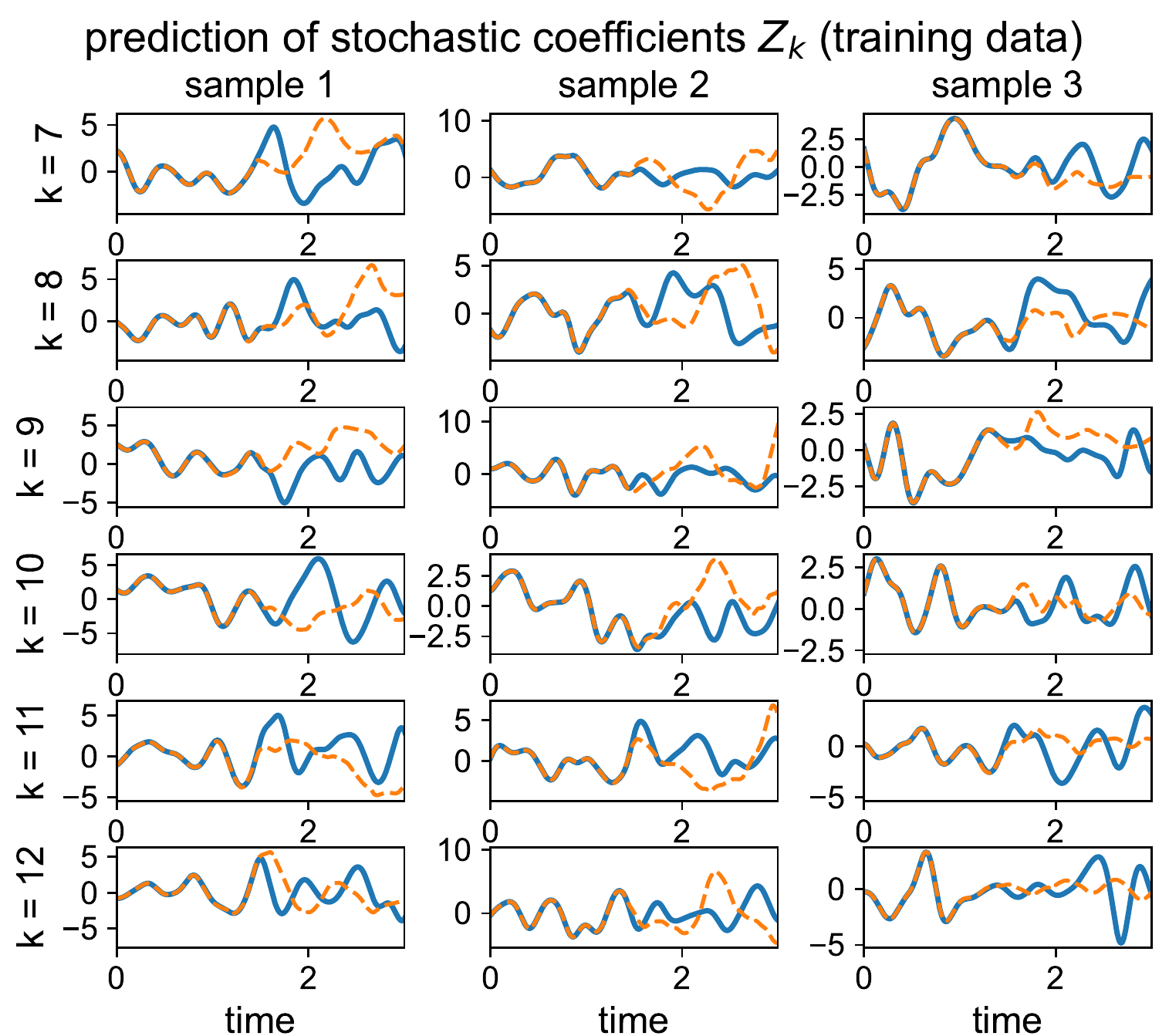}
}\subfloat[trajectory prediction using prediction data]{\includegraphics[scale=0.55]{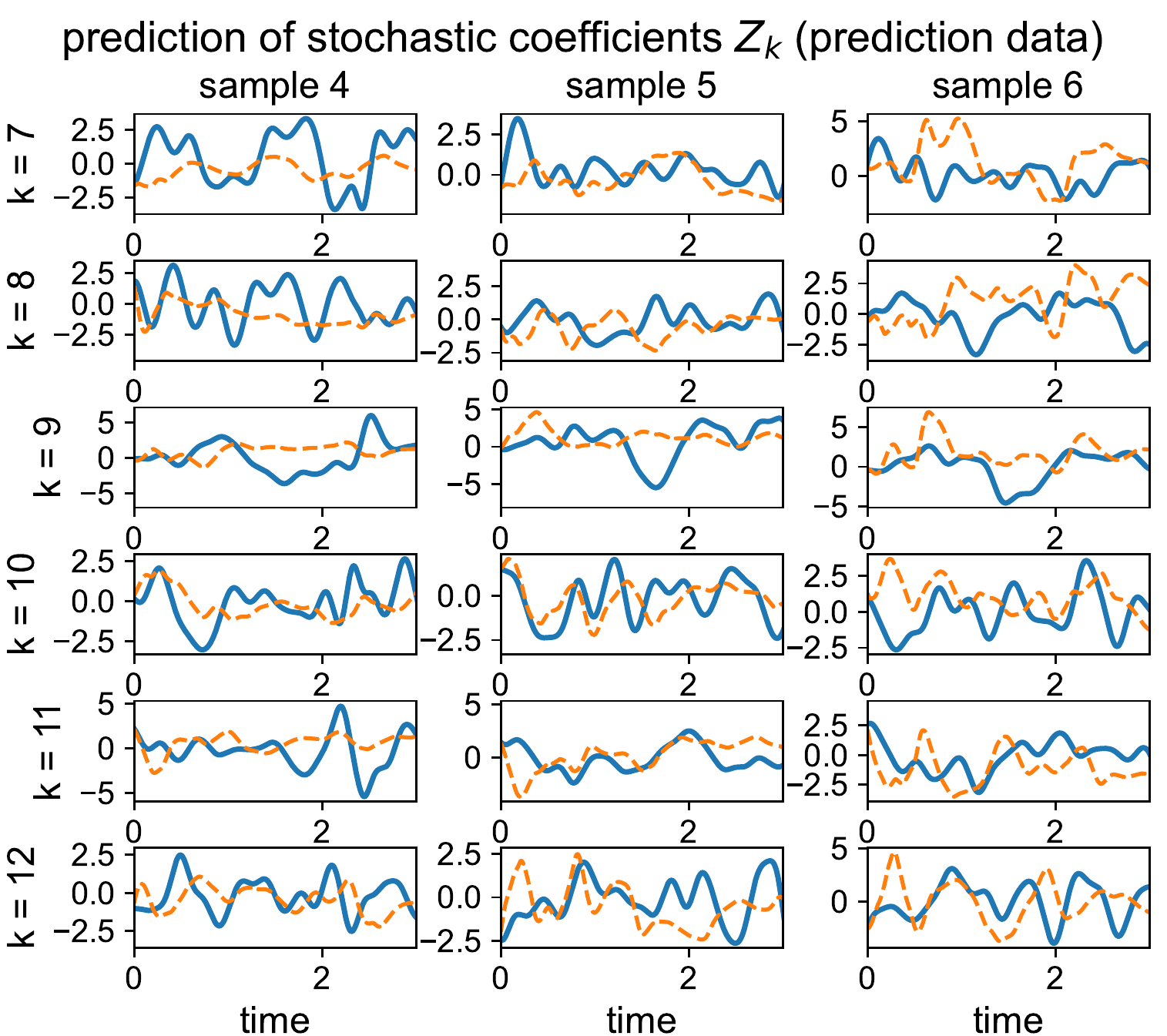}

}

\caption{Training and prediction using the direct stochastic model. The first row shows the training iterations of errors. 
The second row shows predicted time trajectories of stochastic coefficients $Z_{k}$ in
the most energetic modes $k=7,8,9,10,11,12$. Several different sample
trajectories are compared: the 3 samples on the left using the training
data set and the 3 samples on the right using the new prediction data
set. The truth is in solid blue lines while the model prediction is
in dashed orange lines.\label{fig:Predicted-time-trajectories}}

\end{figure}

The training and prediction performance of the direct stochastic model is shown in Figure~\ref{fig:Predicted-time-trajectories}. The loss and mean square errors (MSEs) in the coefficients $Z_k$ and unresolved flux term $\Theta^v_k$ during training iterations are shown in the first row of Figure~\ref{fig:Predicted-time-trajectories}. It appears that the training is effective with the pointwise errors minimized among all the trained samples. Then the trained model is tested on both the previous training data and the new prediction data away from the training set.
The predicted trajectories are recurrently updated in time for a large number of iterations up to a long time $T=3$ (300 iterations compared with only 10 iterations in training). The first 6 most energetic stochastic modes $Z_{k}$ are plotted in the second row of Figure~\ref{fig:Predicted-time-trajectories} in several samples.
Testing on trajectories in the same training set, the predicted solution stays accurate for a while before the solution begins to diverge in time. Referring to the converging rate in Sec.~\ref{subsec:training_data}, the predicted solutions always begin to diverge around the
decorrelation time $T_{\mathrm{decorr}}\sim1.5$ when the autocorrelation decays to zero. This implies the inherent barrier in training
the individual stochastic trajectories beyond the decorrelation time
for a turbulent system containing instability.

More importantly, the trained model fails to predict stochastic trajectories away from the training set. Using new trajectories that are not included in the training data, the prediction diverges immediately and shows no skill in capturing the true trajectory. This shows a typical example of overfitting in training a neural network model. In this specific example, we suspect that the failure can be attributed to combinations of several issues. First, we suspect that the required amount of data to capture the large degrees of uncertainties in this high-dimensional problem is much larger than what we used in this experiment. Second, the stochastic components of the dynamical equations \eqref{partialsystem} are all conditionally unstable modes. With this inherent stability, identification of a stable neural-network modeling becomes a challenging issue, especially if no additional structures are imposed as in our experiments where the standard LSTM model with the residual structure in \eqref{partialsystem} is used. In addition to these issues, the Monte-Carlo error induced by the empirical average in \eqref{R^M_empirical} will amplify difficulties which translates into a numerically expensive training procedure when the true $\mathbf{u}$ is not available in which one has to also learn the unresolved term $\Theta^m$ in \eqref{eq:model_ml_red1}.

This practical issue motivates the idea of fitting response statistics discussed in Section~\ref{empirical_statistical_fitting}, especially when we are mostly interested in the statistical prediction of moments generated by the ensemble averages. As for the instability issue, we consider the use of damping and forcing parameterization discussed in Section~\ref{sec2.2.1} on the neural-network models.

\end{document}